\documentclass[iop,revtex4]{emulateapj}

\usepackage{color}
\usepackage{rotating}

\newcommand{\cht}{CH$_3$C$_2$H(J=12-11)}
\newcommand{\ch}{CH$_3$C$_2$H}
\newcommand{\msun}{M$_{\odot}$}
\newcommand{\lsun}{L$_{\odot}$}

\newcommand{\kms}{km\,s$^{-1}$}       

\newcommand{\cmthree}{cm$^{-3}$}
\newcommand{\um}{$\mu$m}                                 
\newcommand{\amin}{$^{\prime}$}                   
\newcommand{\asec}{$^{\prime \prime}$}
\newcommand{\adeg}{$^{\circ}$}
\newcommand{\higal}{Hi-GAL}
\newcommand{\gapprox}{$\stackrel {>}{_{\sim}}$}   
\newcommand{\lapprox}{$\stackrel {<}{_{\sim}}$}

\shorttitle{Calibration of evolutionary indicators in High-mass star formation}
\shortauthors{Molinari et al.}

\begin{document}


\title{Calibration of evolutionary diagnostics in high-mass star formation.}


\author{S. Molinari}
\affil{Istituto Nazionale di Astrofisica - IAPS, Via Fosso del Cavaliere 100, I-00133 Roma, Italy}
\email{molinari@iaps.inaf.it}

\author{M. Merello}
\affil{Istituto Nazionale di Astrofisica - IAPS, Via Fosso del Cavaliere 100, I-00133 Roma, Italy}

\author{D. Elia}
\affil{Istituto Nazionale di Astrofisica - IAPS, Via Fosso del Cavaliere 100, I-00133 Roma, Italy}

\author{R. Cesaroni}
\affil{Istituto Nazionale di Astrofisica - Osservatorio di Arcetri, Largo E. Fermi 5, I-50125 Firenze, Italy}

\author{L. Testi\altaffilmark{1,2}}
\affil{European Southern Observatory, Garching, Germany}

\and

\author{T. Robitaille}
\affil{Max Planck Institut f\"ur Astronomie, Heidelberg, Germany}


\altaffiltext{1}{Istituto Nazionale di Astrofisica - Osservatorio di Arcetri, Largo E. Fermi 5, Firenze}
\altaffiltext{2}{Excellence Cluster Universe, Boltzmannstr. 2, D-85748 Garching, Germany}


\begin{abstract}
The evolutionary classification of massive clumps that are candidate progenitors of high-mass young stars and clusters relies on a variety of independent diagnostics based on observables from the near-infrared to the radio. 

A promising evolutionary indicator for massive and dense cluster-progenitor clumps is the L/M ratio between the bolometric luminosity and the mass of the clumps. 
With the aim of providing a quantitative calibration for this indicator we used SEPIA/APEX to obtain \cht\ observations, that is an excellent thermometer molecule probing densities $\geq 10^5$\cmthree , toward 51 dense clumps with M$\geq$1000\msun, and uniformly spanning -2 \lapprox\  Log(L/M) [\lsun /\msun] \lapprox\  2.3. 

We identify three distinct ranges of L/M that can be associated to three distinct phases of star formation in massive clumps. For L/M $\leq$1 no clump is detected in \ch , suggesting an inner envelope temperature below $\sim$30K. For 1\lapprox L/M \lapprox\ 10 we detect 58\%\ of the clumps, with a temperature between $\sim$30 and $\sim$35 K independently from the exact value of L/M; such clumps are building up luminosity due to the formation of stars, but no star is yet able to significantly heat the inner clump regions. For L/M\gapprox\ 10 we detect all the clumps, with a gas temperature rising with Log(L/M), marking the appearance of a qualitatively different heating source within the clumps; such values are found towards clumps with UCHII counterparts, suggesting that the quantitative difference in T $vs$ L/M behaviour above L/M$\sim$10 is due to the first appearance of ZAMS stars in the clumps.

\end{abstract}


\section{Introduction}

Molecular clumps, massive (M$\geq$10$^3$\msun ) and cold (T$\leq$30K) condensations of dense gas and dust ($\Sigma \geq 0.1$g cm$^{-2}$) with sizes between 0.1 and few parsecs,  are the primary sites for the formation of stellar clusters. Either by accretion of turbulent cores fragmenting into the clump \citep{mck03} or by a more dynamical process of competitive accretion where initial seeds migrate within the clump hunting for material to accrete \citep{Bonnell+2001}, clump-hosted protoclusters are the sites where intermediate and high-mass stars form (e.g. \citealt{Testi+1999, deWit+2004, Faustini+2009}). 
There is evidence that clumps hosting massive cores in very early stages of formation are already associated with populations of lower mass YSOs visible in the near-IR and hence relatively more evolved (e.g., \citealt{Faustini+2009}). 
Although the early products in clump-hosted protoclusters appear to be relatively low mass YSOs, the appearance of intermediate and high-mass forming protostars is the event that drives the radiative energy budget in these systems. \cite{Molinari+2008} showed that the massive forming objects accreting with a rate proportional to their mass \citep{mck03} show a dramatic increase in radiated bolometric luminosity as a function of time while their core envelope mass decreases only slightly; this can be followed very conveniently in a L$_{bol}$-M$_{env}$ diagram (see Fig. \ref{lm-diag}) where simple-model evolutionary tracks mark the path of star-forming clumps along three basic phases. Initially clumps are  "pre-stellar" or undergoing very low rates of star formation (red dots), with an SED resembling very well a modified blackbody with no  detectable continuum at $\lambda \leq$ 100 \um . In a second phase, that we call "protostellar", the clump starts to exhibit far-infrared continuum radiation well in excess of a single-temperature modified blackbody;
this is interpreted as the indication that ongoing star formation becomes able to significantly heat the clump's interior (the blue dots).
As accretion proceeds the clumps increase their luminosity more and more (following the ascending portion of the model evolutionary tracks in Fig. \ref{lm-diag}) until they reach the luminosity typical for UltraCompact (UC)HII regions 
(see Fig. 9 of \citealt{Molinari+2008}). Subsequent evolution is modelled as envelope dispersal by stellar winds and outflows. Independent evolutionary indications from 
the association with methanol masers or radio continuum \citep{SanchezMonge+2013} or with the properties of ammonia dense gas \citep{Giannetti+2013} are consistent with the predictions in \cite{Molinari+2008}.

With the \higal\ project \citep{Molinari2010b, Molinari+2016a}, 
we detected and characterised several tens of thousands of dense clumps \citep{Elia+2016}, enabling statistical studies mapping star formation history and rates throughout the Milky Way \citep{Veneziani+2013, Veneziani+2016}. It is then important that SFR probes as well as evolutionary tools like the L$_{bol}$-M$_{env}$ diagram are better and better characterised and \emph{calibrated} against indicators that can independently probe the temperatures of the gas in the innermost regions of massive clumps, where the most massive YSOs are forming. Among the best tracers in this respect is \ch\ (methyl-acetylene), that is optically thin in most typical conditions in massive clumps and whose $K$-ladder rotational transitions are collisionally excited at densities in excess of 10$^5$\cmthree (e.g., \citealt{Bergin+1994, Brand+2001, Fontani+2002, TMD2003, Fontani+2004}).

\section{Observations}

\begin{turnpage}
\begin{deluxetable*}{rccccccccccccc}
\tabletypesize{\scriptsize}
\tablecaption{Sources sample and results\label{targetlist}; sources are ordered in increasing L/M.}
\tablecolumns{14}
\tablewidth{0pt}
\tablehead{
\colhead{\#} & \colhead{RA} & \colhead{DEC} & \colhead{Dist} & \colhead{Mass} & \colhead{L/M} & \colhead{r.m.s.} & \multicolumn{4}{c}{CH$_3$C$_2$H (12-11) Integrated Intensity} & \colhead{$\Delta v$} & \colhead{T [CH$_3$C$_2$H]} & \colhead{T$_{dust}$} \\
\colhead{} & \colhead{($^h$\, \amin \, \asec)} & \colhead{(\adeg \, \amin \, \asec)} & \colhead{(kpc)} & \colhead{(10$^4$ \msun)} &  \colhead{(\lsun /\msun)} & \colhead{(mK)\tablenotemark{a} } & \multicolumn{4}{c}{(K $\cdot$ km s$^{-1}$)\tablenotemark{a} } & \colhead{(\kms)} & \colhead{(K)} & \colhead{(K)} \\
\colhead{} & \colhead{} & \colhead{} & \colhead{} & \colhead{} & \colhead{} & \colhead{} & \colhead{(K=0)} & \colhead{(K=1)} & \colhead{(K=2)} & \colhead{(K=3)} & \colhead{} & \colhead{} & \colhead{}
} 
\startdata 
 1   & 15 43 51.9 & -53 58 11.3 & 2.77 & 1.58 & 0.01  &  49 & $-$ & $-$ & $-$ & $-$ &  & & 7.8 \\
 2   & 14 10 11.8 & -61 42 59.3 & 7.76 & 2.87 & 0.01  &  53 & $-$ & $-$ & $-$ & $-$ &  & & 7.7 \\
 3   & 15 32 50.5 & -55 58 08.2 & 10.3 & 21.2 & 0.01  &  50 & $-$ & $-$ & $-$ & $-$ &  & & 7.6 \\
 4   & 16 02 15.1 & -53 21 59.2 & 9.8 & 8.1 & 0.01  &  48 & $-$ & $-$ & $-$ & $-$ &  & & 7.0 \\ 
 5   & 16 27 56.9 & -47 19 45.0 & 5.1 & 0.62 & 0.01  &  53 & $-$ & $-$ & $-$ & $-$ &  & & 7.3  \\
 6   & 14 21 44.0 & -60 52 09.0 & 8.2 & 1.18 & 0.02  &  54 & $-$ & $-$ & $-$ & $-$ &  & & 7.9 \\ 
 7   & 16 55 04.8 & -43 14 34.8 & 5.0 & 1.41 & 0.02  &  49 & $-$ & $-$ & $-$ & $-$ &  & & 8.7 \\ 
 8   & 15 46 26.2 & -53 55 46.9 & 10.7 & 1.78 & 0.03  &  52 & $-$ & $-$ & $-$ & $-$ &  & & 8.2 \\ 
 9   & 16 11 16.4 & -52 02 57.7 & 4.2 & 0.53 & 0.03  &  51 & $-$ & $-$ & $-$ & $-$ &  & & 8.9 \\ 
 10 & 15 56 31.6 & -52 38 03.0 & 8.6 & 1.09 & 0.04  &  47 & $-$ & $-$ & $-$ & $-$ &  & & 8.2 \\ 
 11 & 17 02 43.6 & -42 20 53.7 & 10.6 & 0.89 & 0.05  &  51 & $-$ & $-$ & $-$ & $-$ &  & & 8.8 \\ 
 12 & 16 37 58.7 & -47 08 56.0 & 10.7 & 2.16 & 0.06  &  51 & $-$ & $-$ & $-$ & $-$ &  & & 9.2 \\ 
 13 & 16 34 13.7 & -47 51 18.7 & 8.4 & 1.3 & 0.08  &  51 & $-$ & $-$ & $-$ & $-$ &  & & 9.6 \\ 
 14 & 16 23 26.8 & -49 29 55.5 & 9.8 & 0.86 & 0.10  &  51 & $-$ & $-$ & $-$ & $-$ &  & & 9.2 \\ 
 15 & 14 39 26.0 & -60 01 54.0 & 8.8 & 0.63 & 0.12  &  53 & $-$ & $-$ & $-$ & $-$ &  & & 9.7 \\ 
 16 & 15 16 32.5 & -58 08 31.3 & 9.4 & 0.91 & 0.15  &  51 & $-$ & $-$ & $-$ & $-$ &  & & 9.9 \\ 
 17 & 16 35 27.5 & -47 49 32.8 & 7.2 & 0.58 & 0.18  &  50 & $-$ & $-$ & $-$ & $-$ &  & & 10.9 \\ 
 18 & 15 51 00.9 & -54 26 50.8 & 10.4 & 1.78 & 0.24  &  49 & $-$ & $-$ & $-$ & $-$ &  & & 11.2  \\ 
 19 & 17 14 21.7 & -39 29 21.8 & 10.5 & 0.60 & 0.30  &  53 & $-$ & $-$ & $-$ & $-$ &  & & 10.9 \\ 
 20 & 16 12 38.5 & -51 37 32.8 & 9.86 & 0.52 & 0.37  &  53 & $-$ & $-$ & $-$ & $-$ &  & & 12.1 \\ 
 21 & 13 12 22.8 & -62 34 59.4 & 5.9 & 2.0 & 0.49  &  50 & $-$ & $-$ & $-$ & $-$ &  & & 12.2 \\ 
 22 & 15 14 32.2 & -58 11 22.2 & 8.24 & 0.60 & 0.59  &  51 & $-$ & $-$ & $-$ & $-$ &  & & 13.6 \\ 
 23 & 16 10 19.7 & -51 49 44.9 & 9.3 & 0.60 & 0.73  &  46 & $-$ & $-$ & $-$ & $-$ &  & & 14.1 \\ 
 24 & 11 15 12.3 & -61 18 52.0 & 8.12 & 0.83 & 1.0  &  14 & 0.115$\pm$0.015 & 0.106$\pm$0.015 & 0.053$\pm$0.016 & 0.035$\pm$0.016 & 4.19 & 36.1$\pm$4.4 & 13.8 \\ 
 25 & 13 11 17.1 & -62 46 38.5 & 6.94 & 0.59 & 1.1  &  53 & 0.458$\pm$0.041 & 0.380$\pm$0.038 & 0.134$\pm$0.040 & 0.145$\pm$0.038 & 2.67 & 34.4$\pm$4.5 & 15.9 \\ 
 26 & 16 34 56.4 & -47 34 37.2 & 10.8 & 0.70 & 1.7  &  50 & 0.205$\pm$0.048 & 0.175$\pm$0.052 & 0.083$\pm$0.041 & 0.063$\pm$0.041 & 3.41 & 34.9$\pm$1.6 & 16.5 \\ 
 27 & 17 17 32.7 & -37 39 31.8 & 11.2 & 0.72 & 2.0  &  53 & $-$ & $-$ & $-$ & $-$ &  & & 14.8 \\ 
 28 & 15 32 14.4 & -55 52 31.6 & 10.3 & 0.63 & 2.4  &  40 & $-$ & $-$ & $-$ & $-$ &  & & 13.7 \\ 
 29 & 17 11 25.9 & -39 09 12.0 & 5.76 & 0.58 & 2.9  &  53 & 0.533$\pm$0.044 & 0.522$\pm$0.044 & 0.218$\pm$0.041 & 0.168$\pm$0.040 & 3.1 & 34.9$\pm$5.4 & 18.9 \\ 
 30 & 16 41 43.3 & -46 18 39.8 & 9.9 & 0.67 & 3.3  &  49 & $-$ & $-$ & $-$ & $-$ & & & 16.8 \\ 
 31 & 13 12 48.4 & -62 36 15.7 & 6.76 & 0.18 & 4.6  &  35 & $-$ & $-$ & $-$ & $-$ & & & 15.2 \\ 
 32 & 16 12 07.1 & -51 58 30.9 & 10.5 & 1.93 & 5.6  &  54 & $-$\tablenotemark{b} & $-$\tablenotemark{b} & $-$ & $-$ & & & 16.1 \\ 
 33 & 16 12 02.0 & -52 00 53.2 & 10.5 & 1.55 & 7.1  &  54 & 0.353$\pm$0.037 & 0.356$\pm$0.038 & 0.127$\pm$0.034 & 0.094$\pm$0.034 & 2.21 & 31.8$\pm$6.7 & 17.4 \\ 
 34 & 14 03 35.4 & -61 18 23.8 & 6.2 & 0.17 & 7.9  &  36 & $-$ & $-$ & $-$ & $-$ & & & 16.7 \\ 
 35 & 13 11 48.0 & -62 46 39.5 & 6.9 & 0.11 & 8.4  &  20 & $-$\tablenotemark{b} & $-$\tablenotemark{b} & $-$ & $-$ & & & 20.9 \\ 
 36 & 14 26 08.3 & -60 40 21.4 & 7.9 & 0.19 & 10.3 &  24 & $-$ & $-$ & $-$ & $-$ & & & 17.9 \\ 
 37 & 12 03 16.6 & -63 11 17.7 & 11.1 & 0.59 & 11.5 &  22 & 0.350$\pm$0.020 & 0.136$\pm$0.014 & 0.106$\pm$0.014 & 0.100$\pm$0.014 & 2.88 & 36.8$\pm$3.3 & 21.6 \\ 
 38 & 16 12 07.3 & -51 30 01.1 & 9.5 & 0.57 & 12.8 &  45 & $-$\tablenotemark{b} & $-$\tablenotemark{b} & $-$ & $-$ & & & 24.4 \\ 
 39 & 13 11 14.5 & -62 47 25.6 & 6.94 & 0.14 & 19.8 &  34 & 0.575$\pm$0.030 & 0.567$\pm$0.031 & 0.248$\pm$0.029 & 0.243$\pm$0.029 & 2.96 & 40.9$\pm$6.1 & 27.9 \\ 
 40 & 16 48 05.2 & -45 05 07.1 & 9.3 & 0.59 & 21.3 &  53 & 0.575$\pm$0.038 & 0.514$\pm$0.040 & 0.241$\pm$0.037 & 0.164$\pm$0.037 & 3.01 & 34.2$\pm$2.9 & 22.9 \\ 
 41 & 15 00 55.1 & -58 58 50.4 & 10.4 & 0.62 & 22.8 &  53 & 1.020$\pm$0.045 & 0.768$\pm$0.042 & 0.440$\pm$0.041 & 0.281$\pm$0.048 & 2.49 & 34.4$\pm$2.1 & 32.0 \\ 
 42 & 16 01 01.7 & -52 38 51.2 & 8.27 & 0.17 & 23.6 &  35 & 0.192$\pm$0.025 & 0.180$\pm$0.026 & 0.068$\pm$0.023 & 0.059$\pm$0.023 & 1.98 & 33.6$\pm$5.0 & 20.0 \\ 
 43 & 11 15 10.6 & -61 20 33.4 & 8.12 & 0.18 & 25.1 &  20 & 0.730$\pm$0.022 & 0.653$\pm$0.020 & 0.355$\pm$0.200 & 0.240$\pm$0.019 & 3.47 & 37.7$\pm$2.7 & 31.8 \\ 
 44 & 16 36 43.1 & -47 31 22.9 & 10.9 & 1.41 & 28.3 &  51 & 4.687$\pm$0.049 & 4.043$\pm$0.048 & 2.500$\pm$0.046 & 1.964$\pm$0.048 & 3.85 & 43.8$\pm$1.6 & 24.5 \\ 
 45 & 16 35 33.9 & -47 31 10.7 & 10.9 & 0.85 & 28.9 &  52 & 1.181$\pm$0.038 & 1.098$\pm$0.038 & 0.311$\pm$0.037 & 0.460$\pm$0.037 & 3.67 & 38.4$\pm$7.9 & 36.3 \\ 
 46 & 15 09 52.4 & -58 25 33.8 & 8.24 & 0.24 & 36.5 &  36 & 1.368$\pm$0.035 & 1.192$\pm$0.033 & 0.790$\pm$0.034 & 0.412$\pm$0.034 & 3.62 & 40.3$\pm$6.5 & 36.9 \\ 
 47 & 13 32 35.5 & -62 45 29.5 & 7.2 & 0.17 & 38.1 &  35 & 0.275$\pm$0.023 & 0.221$\pm$0.022 & 0.102$\pm$0.022 & 0.125$\pm$0.022 & 1.73 & 43.9$\pm$5.7 & 36.9 \\ 
 48 & 16 21 32.4 & -50 26 47.5 & 3.55 & 0.24 & 52.1 &  35 & 2.870$\pm$0.034 & 2.483$\pm$0.033 & 1.405$\pm$0.031 & 1.103$\pm$0.033 & 3.93 & 40.9$\pm$1.2 & 25.2 \\
 49 & 12 09 58.0 & -62 49 35.8 & 11.35 & 0.86 & 72.0    &  15 & 0.176$\pm$0.014 & 0.136$\pm$0.014 & 0.106$\pm$0.012 & 0.100$\pm$0.014 & 3.52 & 56.5$\pm$6.1 & 21.0 \\ 
 50 & 16 20 11.4 & -50 53 12.2 & 3.85 & 0.46 & 77.0 &  38 & 7.698$\pm$0.008 & 7.043$\pm$0.041 & 4.163$\pm$0.042 & 3.878$\pm$0.001 & 4.98 & 49.2$\pm$0.3 & 24.0 \\ 
 51 & 15 54 06.4 & -53 11 37.8 & 8.44 & 0.55 & 185 &  52 & 1.308$\pm$0.049 & 1.321$\pm$0.049 & 0.682$\pm$0.049 & 0.886$\pm$0.050 & 4.5 & 58.9$\pm$9.4 & 32.9 \\ 
\enddata
\tablenotetext{a}{Main-beam temperatures, assuming $\eta _{mb}$=0.8; reported r.m.s. is for 0.112 \kms\ channels. }
\tablenotetext{b}{K=0,1 lines are detected but no convergence was found in the simultaneous K=0,1,2,3 line fitting.}
\end{deluxetable*}
\end{turnpage}

\subsection{Target selection}
Dense and massive clumps in active phase of star formation were selected from the Hi-GAL survey \citep{Molinari2010b, Molinari+2016a, Elia+2016}. A first list is established by selecting Hi-GAL sources with counterparts in at least four adjacent Herschel bands including the 70\um\ band, and with an available determination of a heliocentric distance \citep{russeil:2011}. The submillimeter portion of the continuum SED in these clumps is fitted with a modified greybody whose temperature represents the bulk of the cold dust in the clumps, and that provides negligible contribution for $\lambda \leq$100\um; the presence of a 70\um\ counterpart is therefore in excess of this cold clump emission and is generally assumed as an indication of ongoing star formation inside the clumps.

We restricted to objects in the longitude range  $-10$\adeg $\leq l \leq -60$\adeg\ and with a 350 \um\ flux $\geq$ 10Jy to be in the best detection conditions irrespectively of the target distance. We then selected the clumps with M$\geq$1000 M$_{\odot}$, that can generate intermediate and high-mass protostars, and with distance \lapprox\ 10 kpc. The resulting sample of objects, conveniently distributed on the far tip of the Galactic Bar, the Norma, Scutum and Sagittarius arms, was finally skimmed down to 51 sources by sampling the distribution of the L/M values to uniformly cover (in Log values) the range of L/M from the initial protostellar phase (L/M$\sim 0.01$, albeit with a 70 \um\ counterpart) to the UCHII phase with L/M \gapprox 100 (hereinafter in units of \lsun / \msun). Target positions and properties from \cite{Elia+2016} are reported in Table \ref{targetlist}; sources distribution in L/M is uniform as a function of distance so that no distance bias is present in the target sample.

\subsection{Data analysis}
Observations of the \cht\ lines ladder (K=0, 3) at $\nu$=205.08073-205.01811~GHz  toward the target clumps were carried out using the new SEPIA receiver equipped with a Band 5 ALMA pre-production cartridge \citep{SEPIA2012} at the ESO Atacama Pathfinder EXperiment telescope, with a main beam HPBW$\sim$30\asec. Data were acquired in ON-OFF mode toward the 70 \um\ peak position of each clump; thanks to values of 0.4-0.45 mm of precipitable water vapour, we could reach typical r.m.s. values in 0.112 \kms\ channels between 50 and 60 mK (in T$_{MB}$, see below) in about 12 minutes integration time on average. More time could be spent on a few sources, pushing the r.m.s. down to values between 20 and 40mK and hence improving the S/N of the detected lines. The CLASS software package was used to accumulate spectra and subtract a first-order baseline. Antenna temperatures were converted into main-beam temperatures using an efficiency of $\eta _{mb}$=0.8 (de Breuck, priv. comm.).
The spectra obtained for the detected sources are reported in Fig.\ref{spectra}, with the location of the four $K$-ladder components marked by vertical ticks. Simultaneous 4-component Gaussian fitting was performed in CLASS, fixing the frequency separation among the different $K$ components to the laboratory values and forcing their linewidths to be all equal. In some cases, sources with marginal detections of the higher $K$ components did not yield convergence in the fitting and they were neglected in subsequent analysis (see Table \ref{targetlist}).

Integrated line fluxes obtained from the Gaussian fitting are reported in table \ref{targetlist}, and were used to compute Boltzmann plots (i.e. the rotational diagram) relating the column density of the emitters normalised by the level degeneracy and the energy of the levels. Using standard procedures (e.g. \citealt{Fontani+2002}) assuming LTE and optically thin conditions that are appropriate for this line and class of sources (e.g., \citealt{Bergin+1994}) the temperature and total column density of \ch\ can be derived from a linear fit to the rotational diagram \citep{GL99}.
Because our observations are single-pointed toward the 70 \um\ peak of the sources we could only derive beam-averaged column densities and we choose to focus on the analysis of the \ch\ rotational temperatures (see Table \ref{targetlist}).

\begin{figure*}
\includegraphics[trim=1.5cm 1.5cm 1.5cm 1.5cm, width=0.33\textwidth]{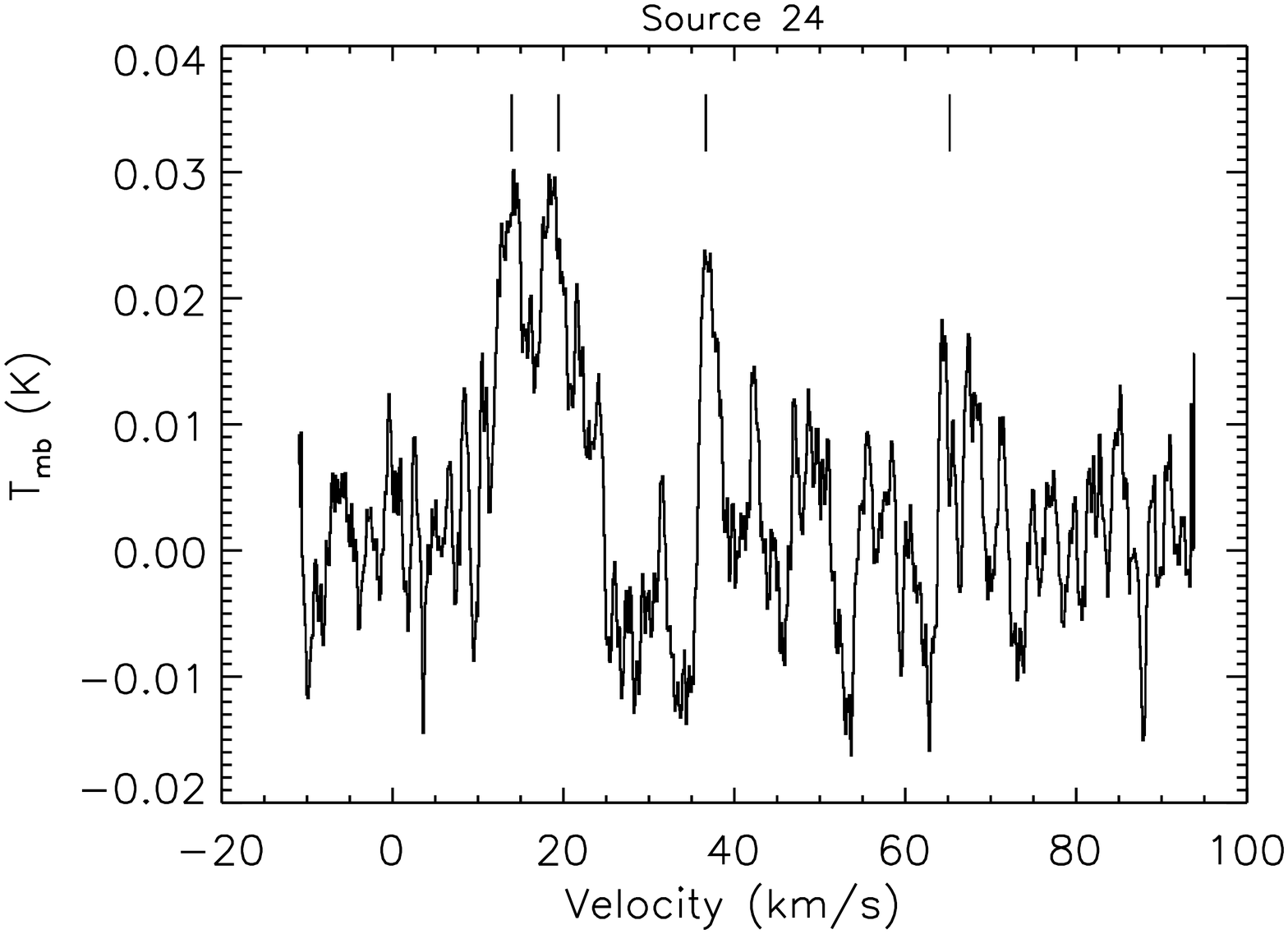}
\includegraphics[trim=1.5cm 1.5cm 1.5cm 1.5cm, width=0.33\textwidth]{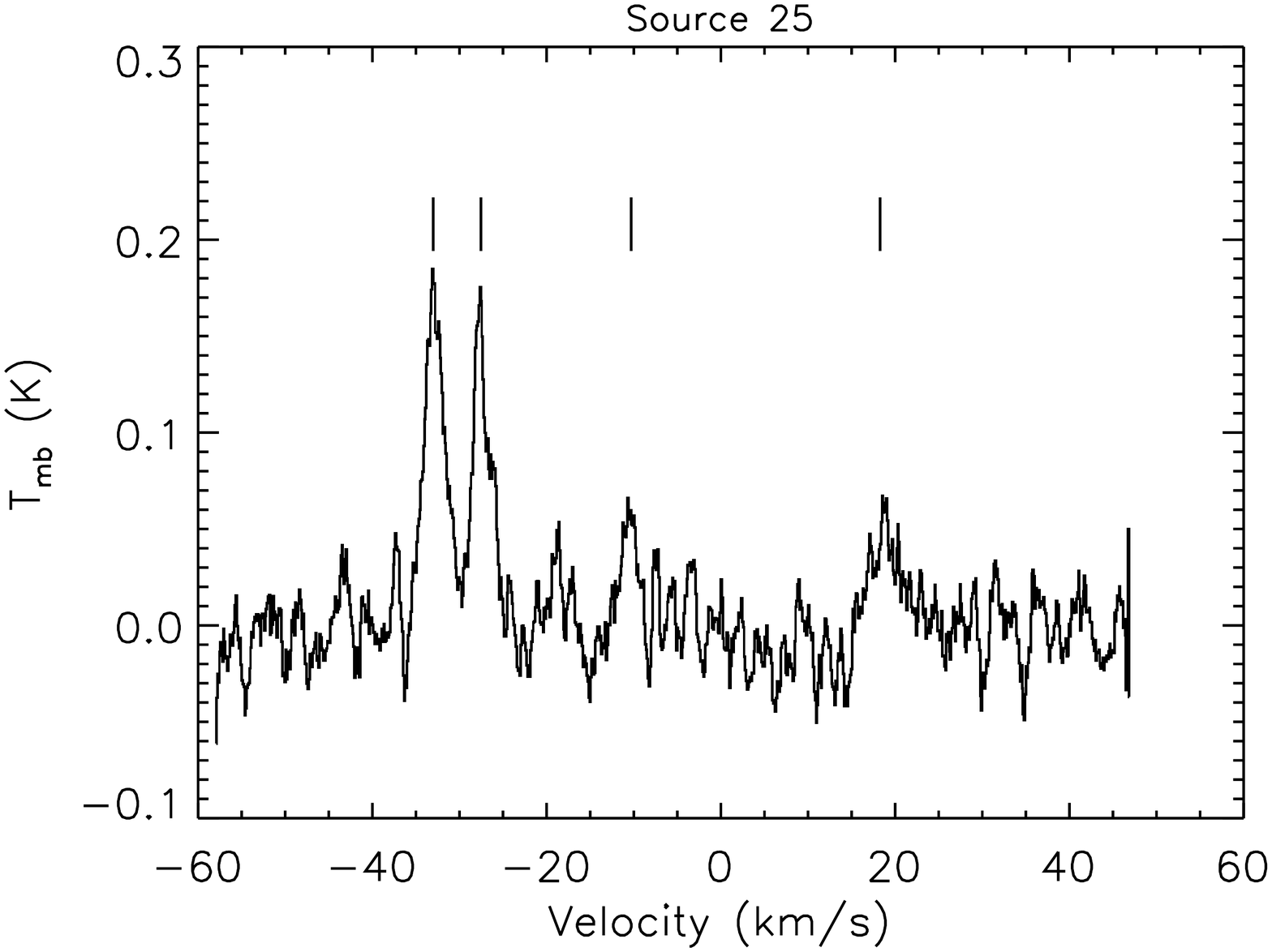}
\includegraphics[trim=1.5cm 1.5cm 1.5cm 1.5cm, width=0.33\textwidth]{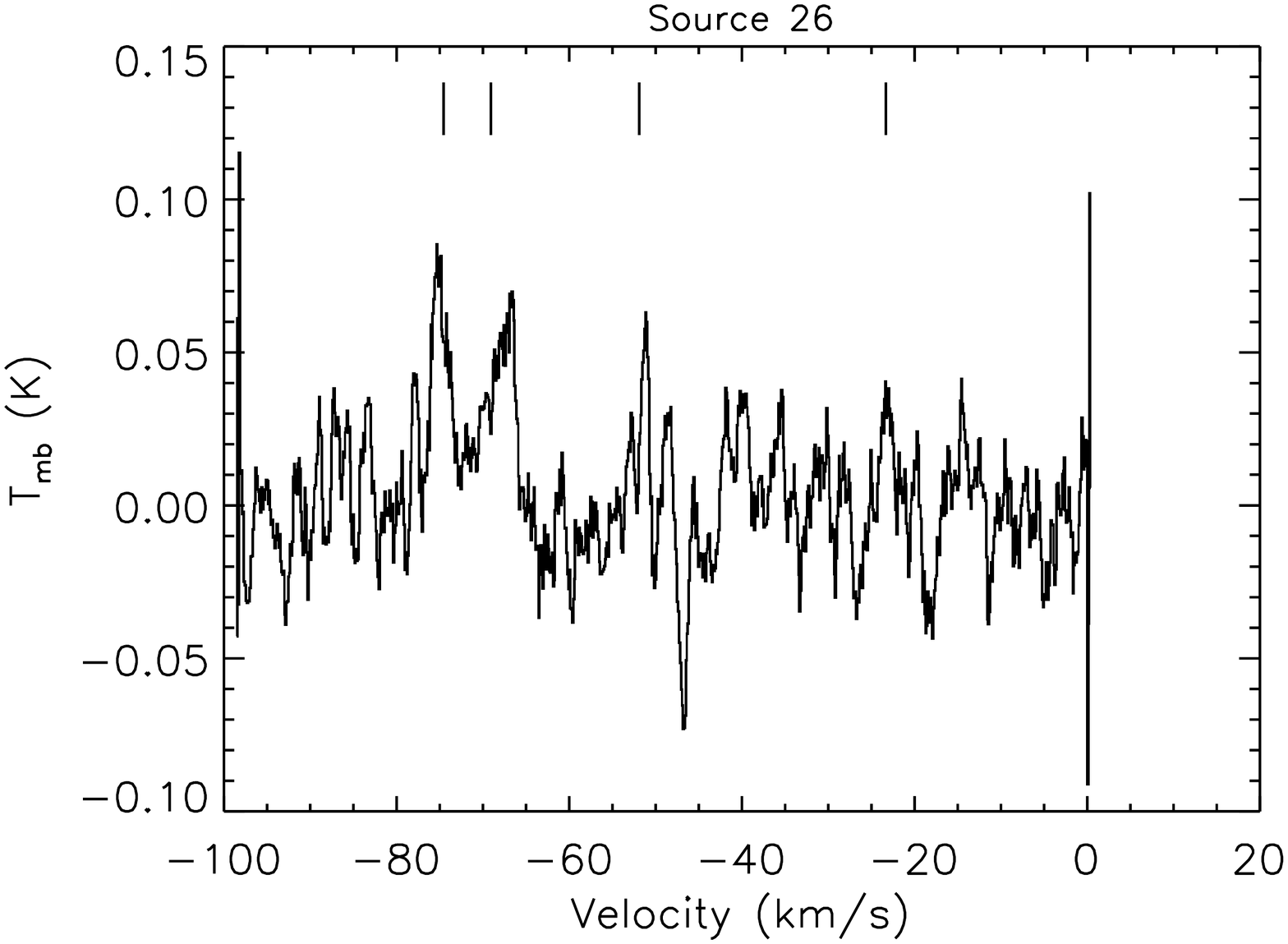}
\includegraphics[trim=1.5cm 1.5cm 1.5cm 1.5cm, width=0.33\textwidth]{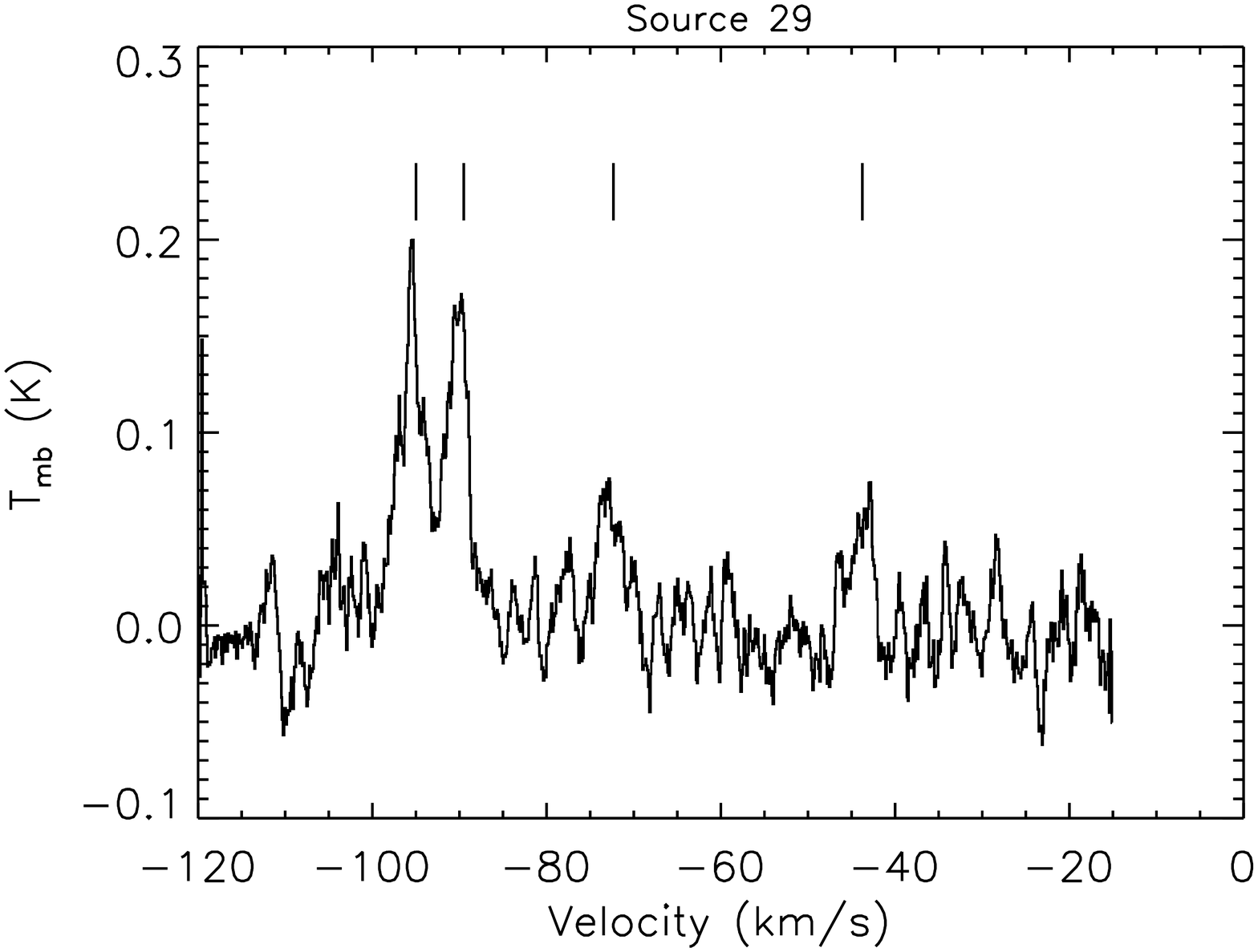}
\includegraphics[trim=1.5cm 1.5cm 1.5cm 1.5cm, width=0.33\textwidth]{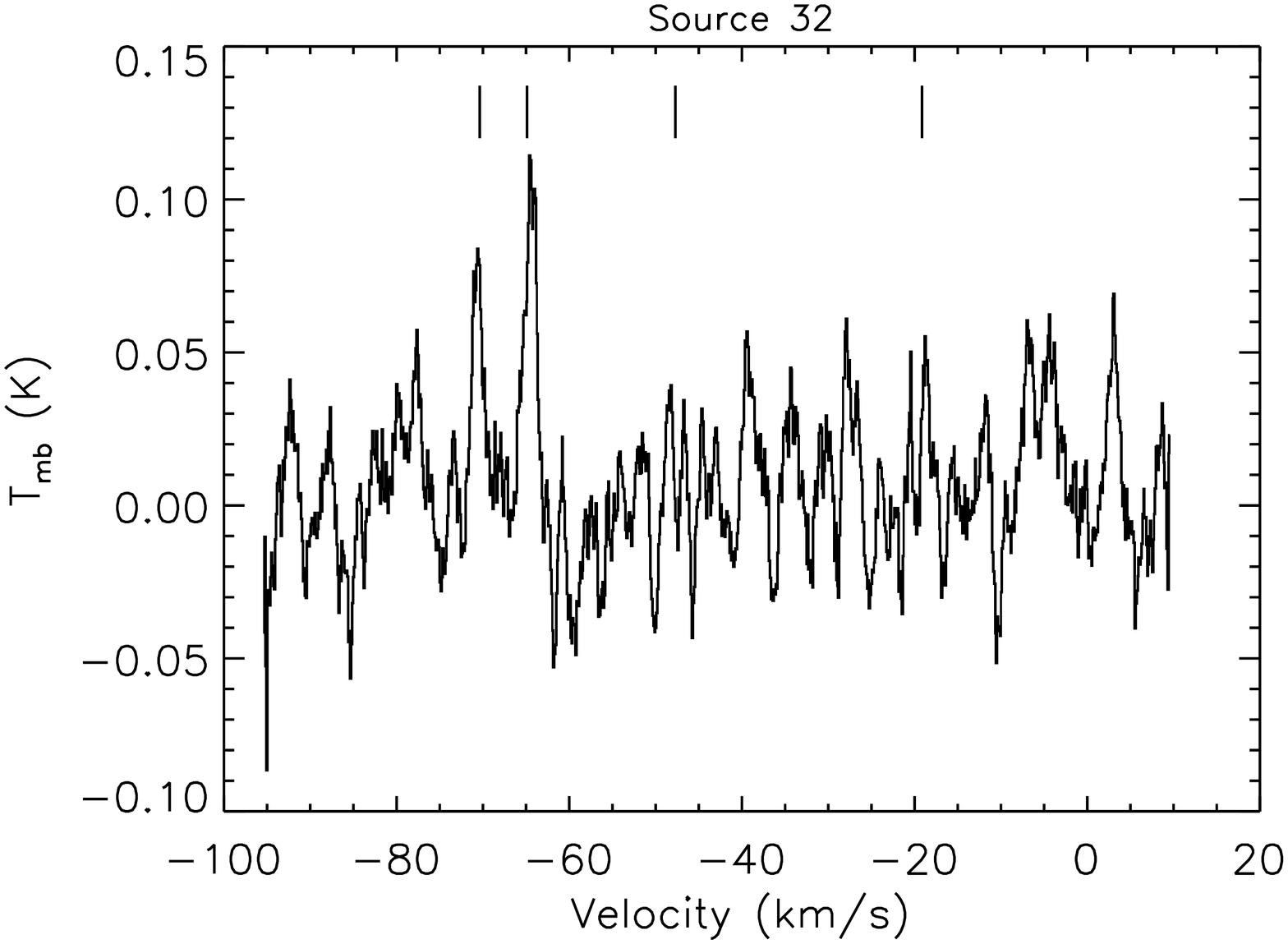}
\includegraphics[trim=1.5cm 1.5cm 1.5cm 1.5cm, width=0.33\textwidth]{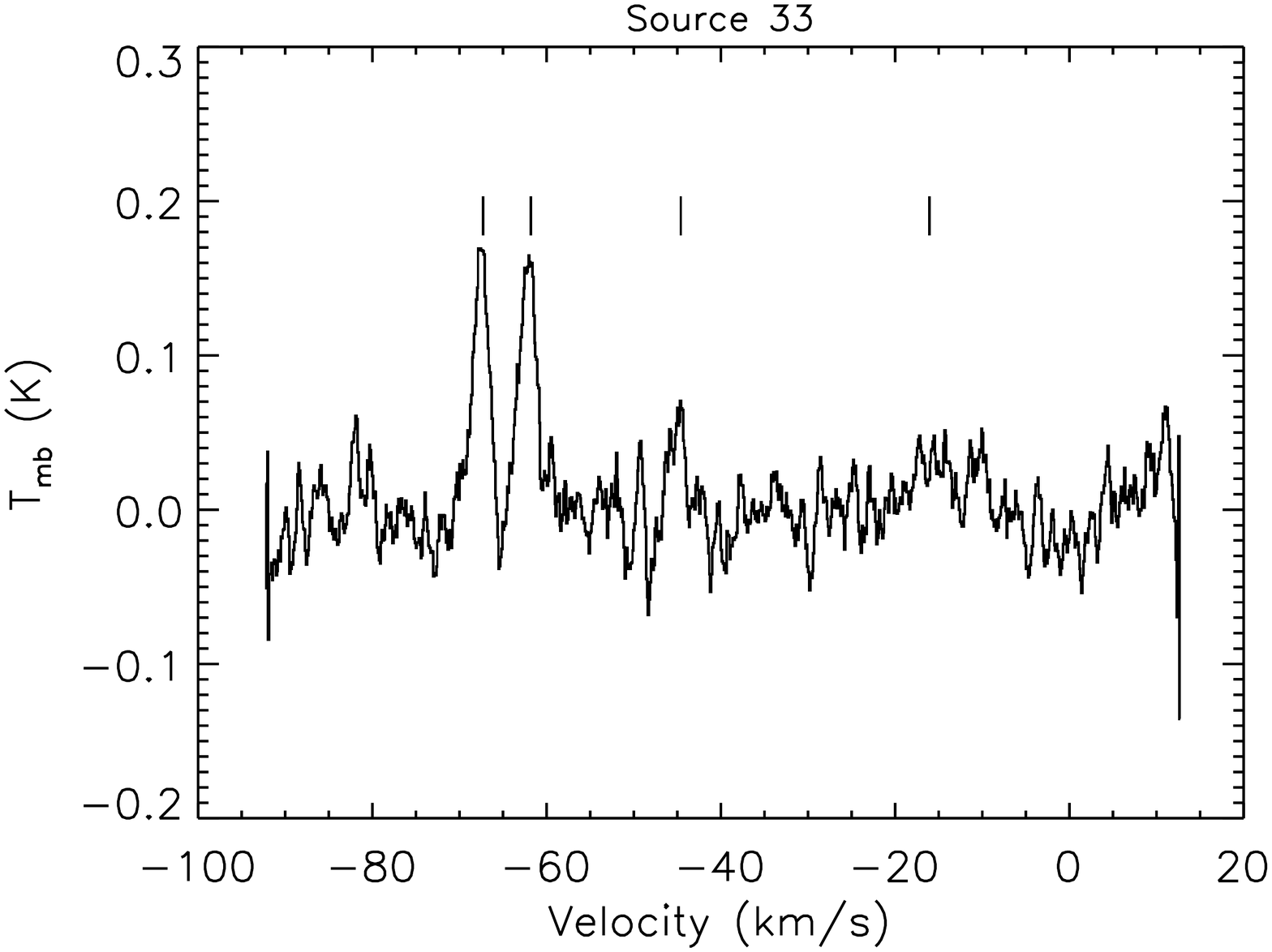}
\includegraphics[trim=1.5cm 1.5cm 1.5cm 1.5cm, width=0.33\textwidth]{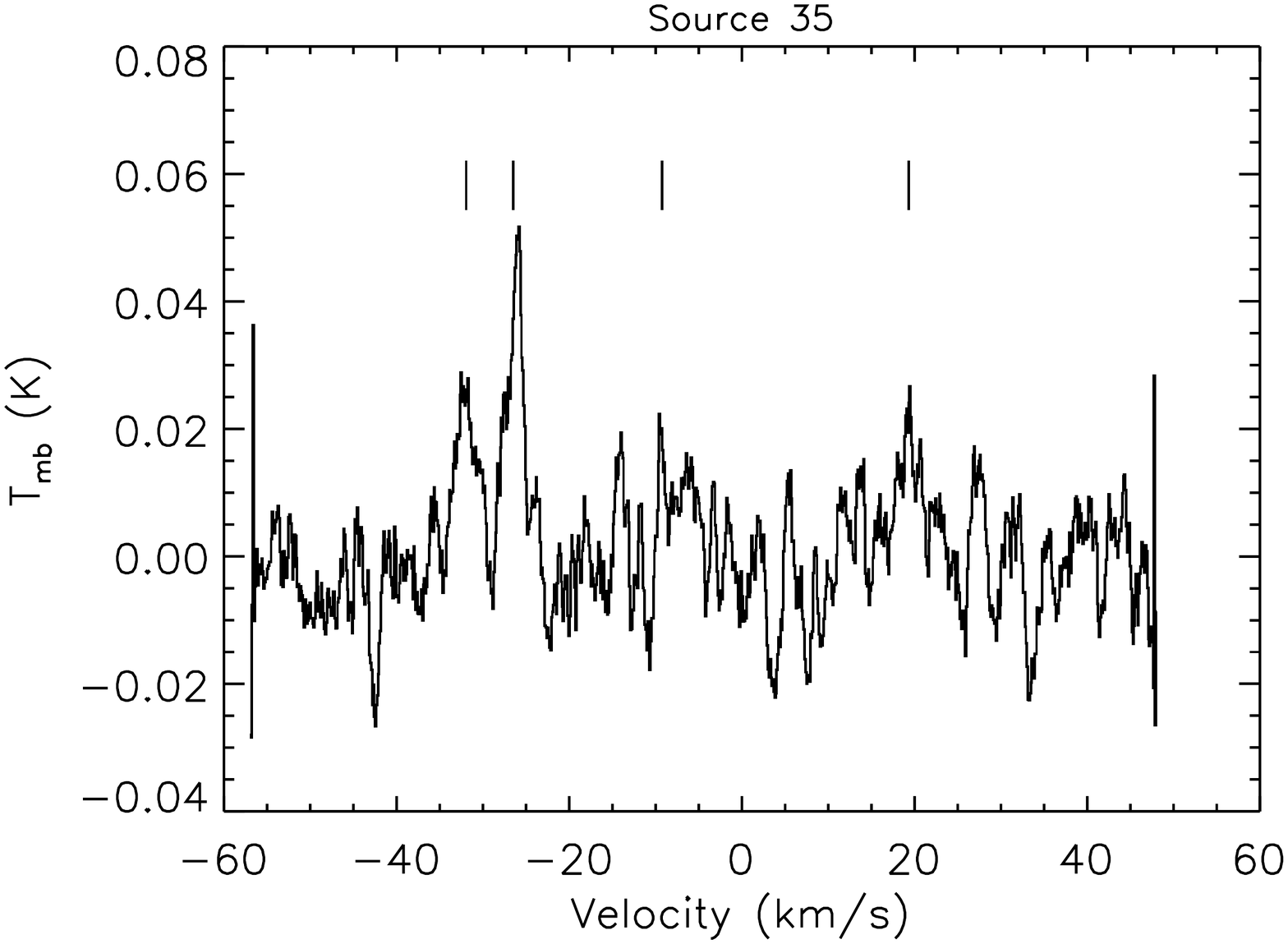}
\includegraphics[trim=1.5cm 1.5cm 1.5cm 1.5cm, width=0.33\textwidth]{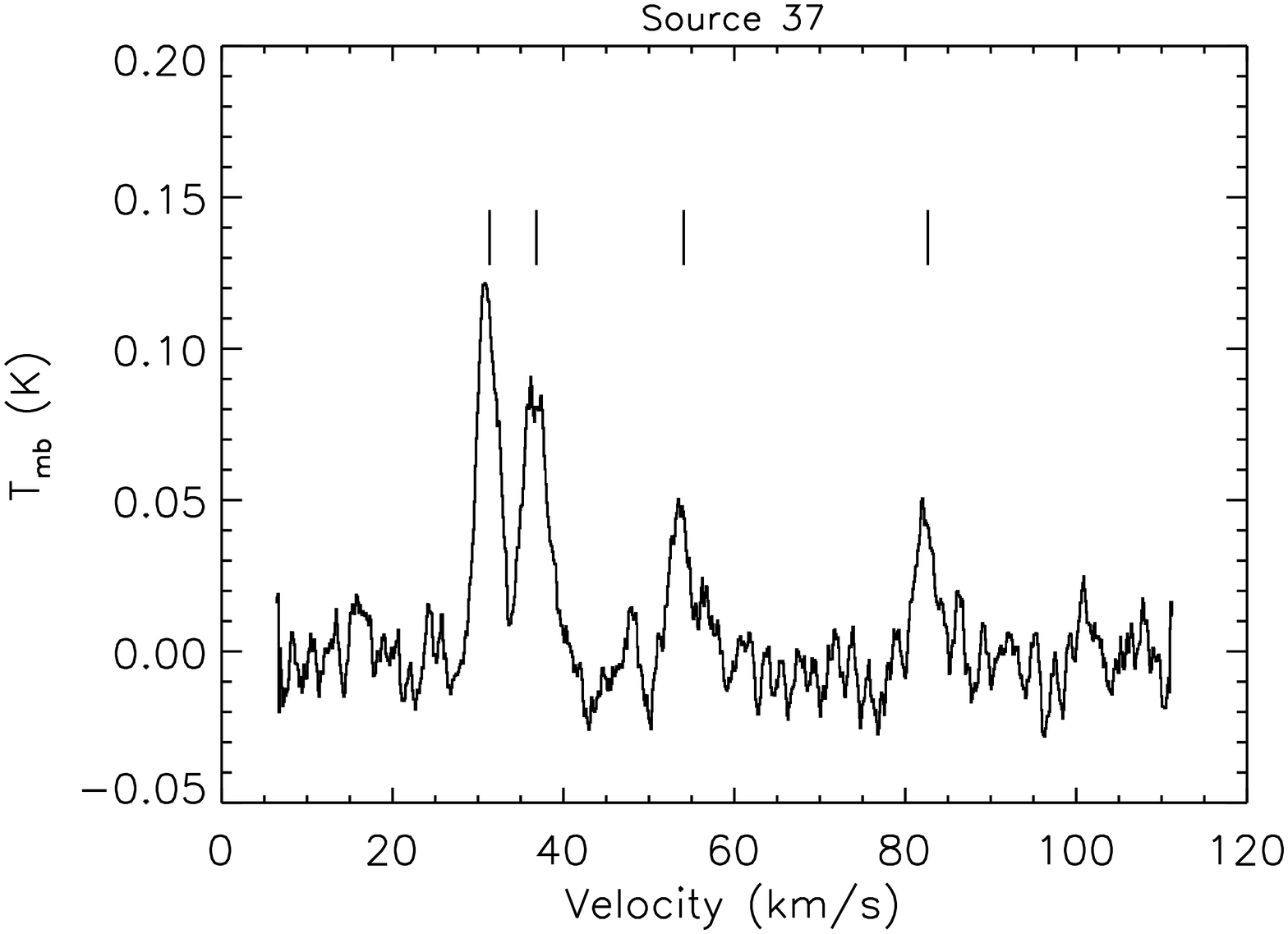}
\includegraphics[trim=1.5cm 1.5cm 1.5cm 1.5cm, width=0.33\textwidth]{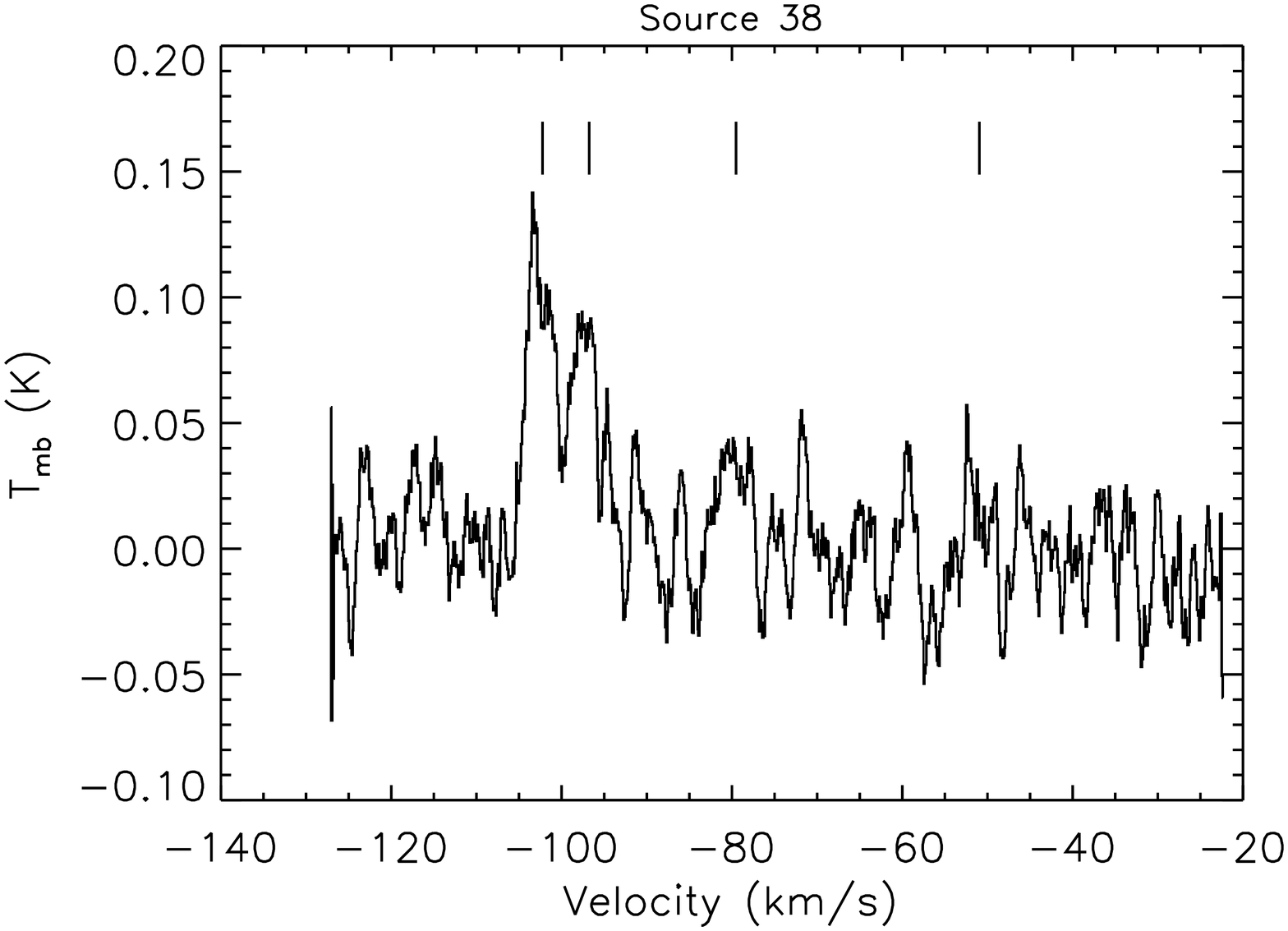}
\includegraphics[trim=1.5cm 1.5cm 1.5cm 1.5cm, width=0.33\textwidth]{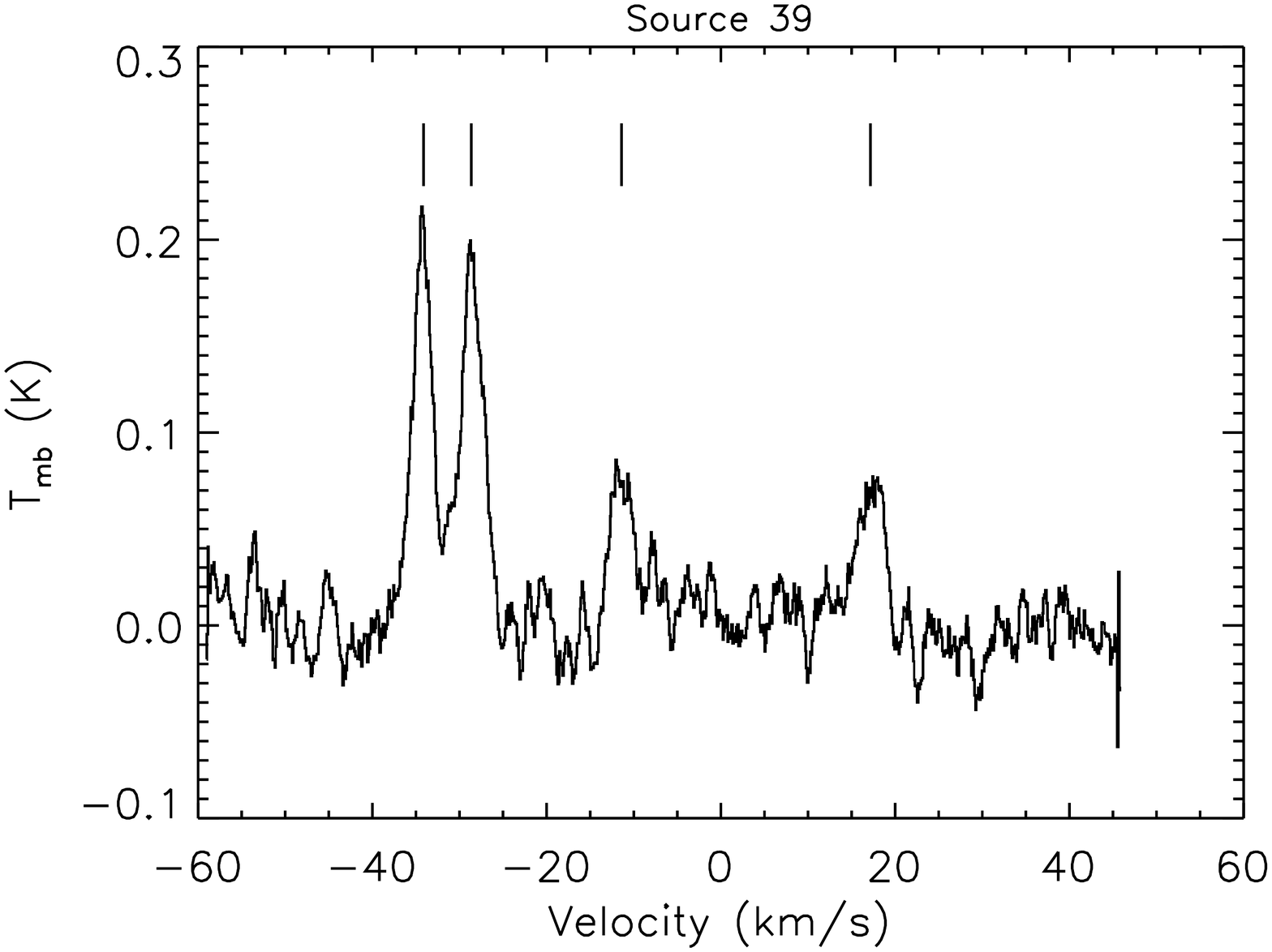}
\includegraphics[trim=1.5cm 1.5cm 1.5cm 1.5cm, width=0.33\textwidth]{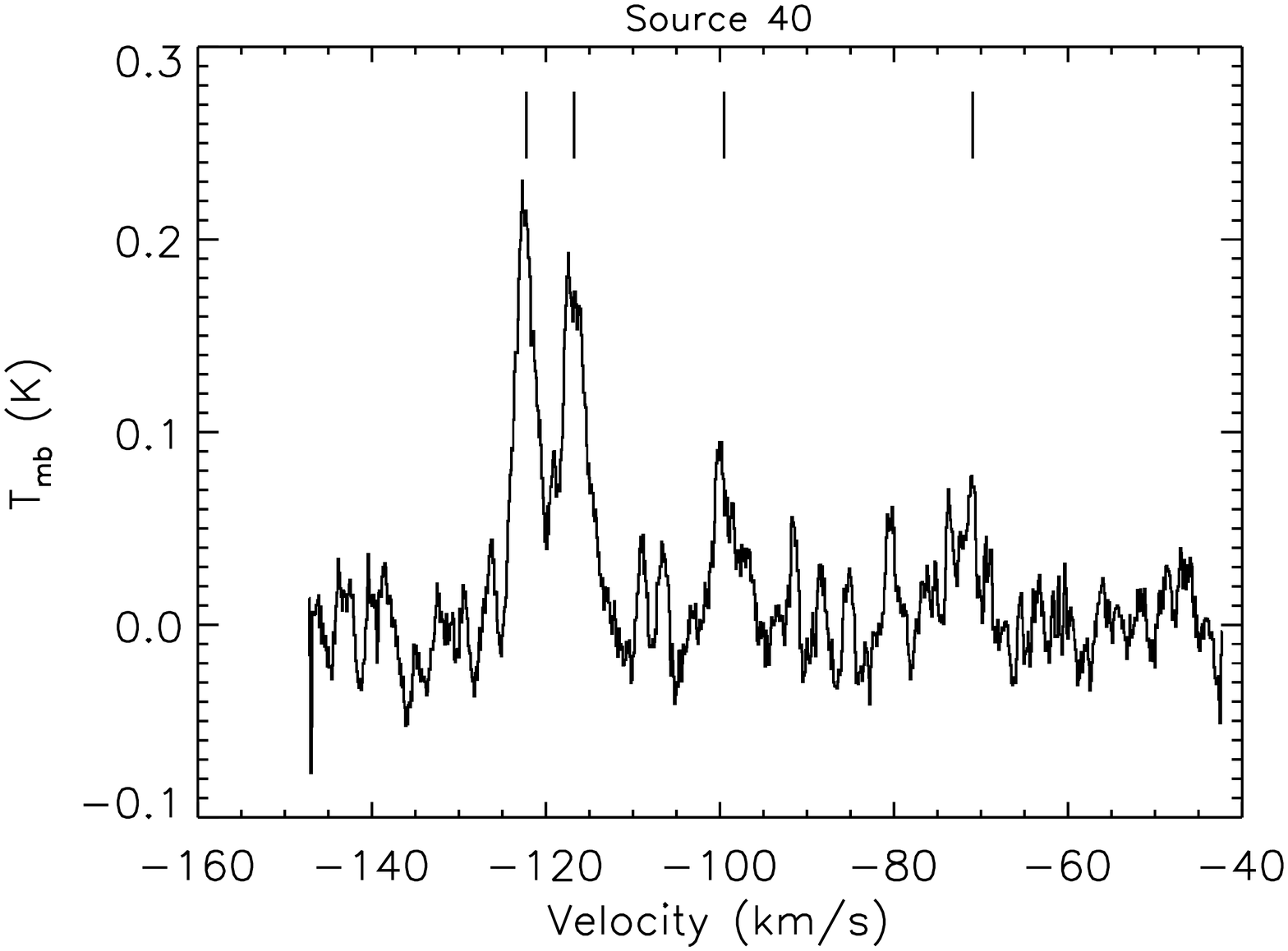}
\includegraphics[trim=1.5cm 1.5cm 1.5cm 1.5cm, width=0.33\textwidth]{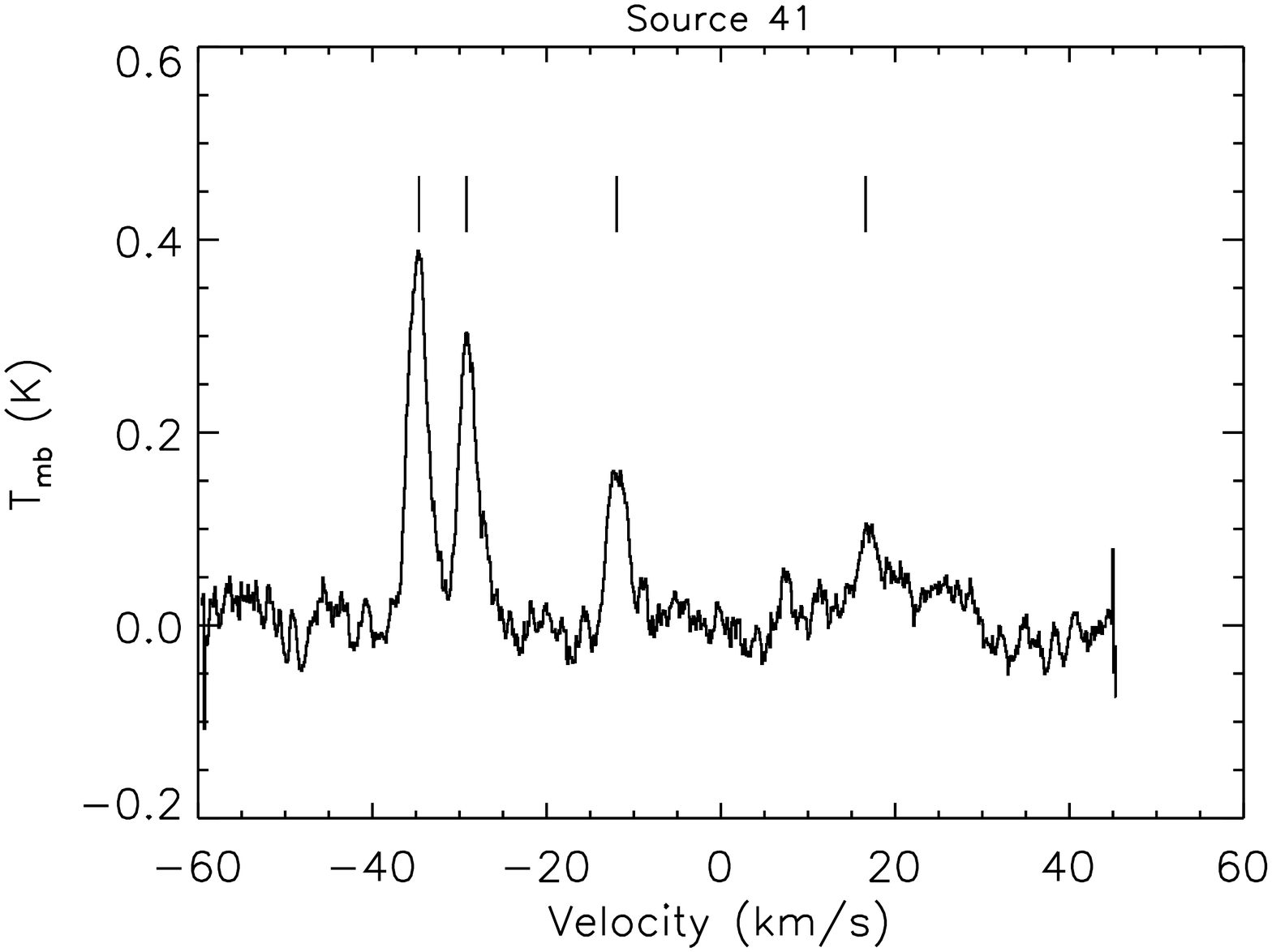}
\caption{CH$_3$C$_2$H(12-11) line spectra for sources with a positive detection; the velocities for the K=0,1,2,3 components are marked with vertical bars. Subsequent panels show the rest of the sources detected; in the case of sources 32, 35 and 38 the K=0,1 lines were detected, but no valid solution was found for the simultaneous Gaussian fitting of all four components. \label{spectra}}
\end{figure*}




\addtocounter{figure}{-1}
\begin{figure*}
\includegraphics[trim=1.5cm 1.5cm 1.5cm 1.5cm, width=0.33\textwidth]{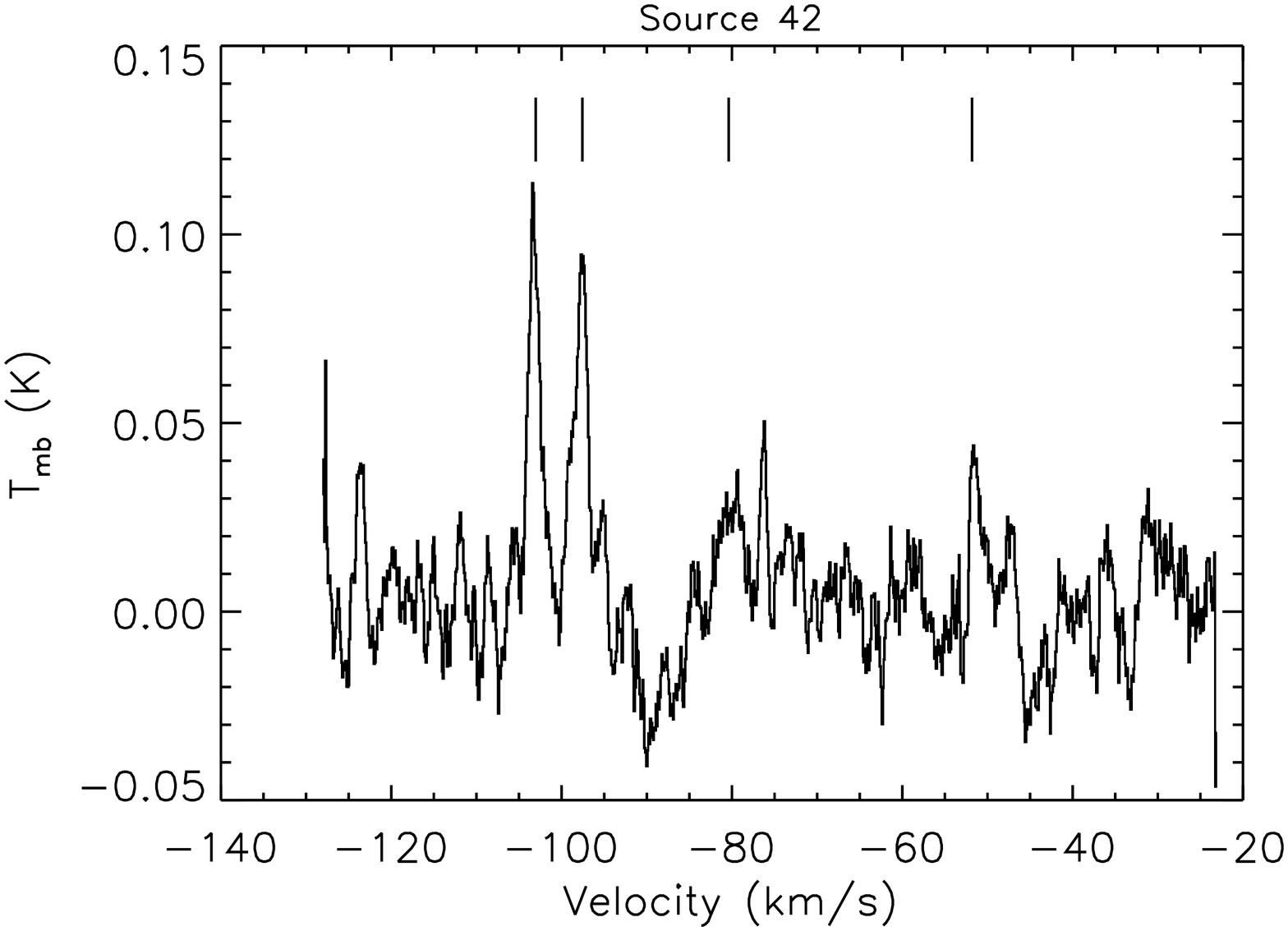}
\includegraphics[trim=1.5cm 1.5cm 1.5cm 1.5cm, width=0.33\textwidth]{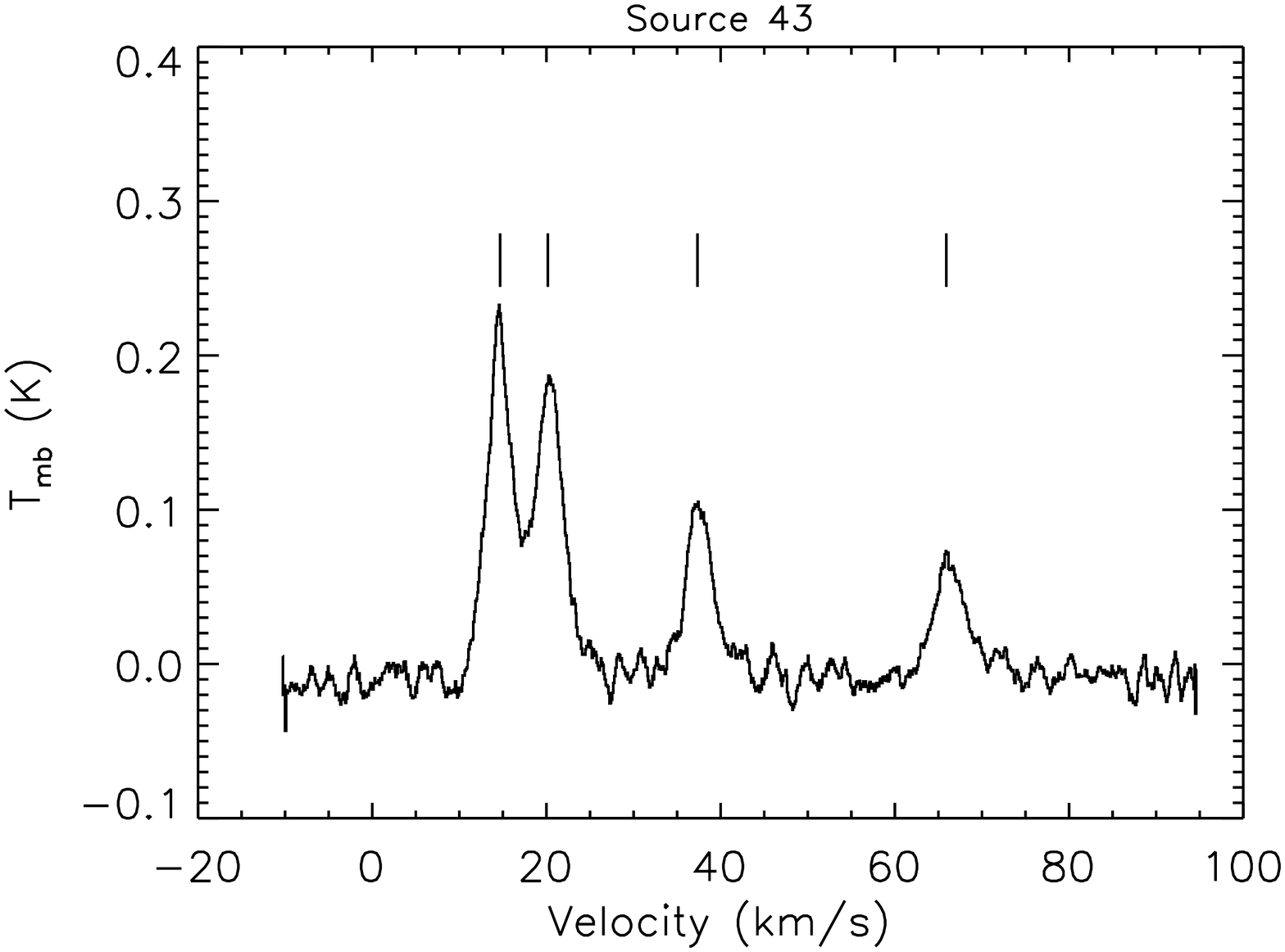}
\includegraphics[trim=1.5cm 1.5cm 1.5cm 1.5cm, width=0.33\textwidth]{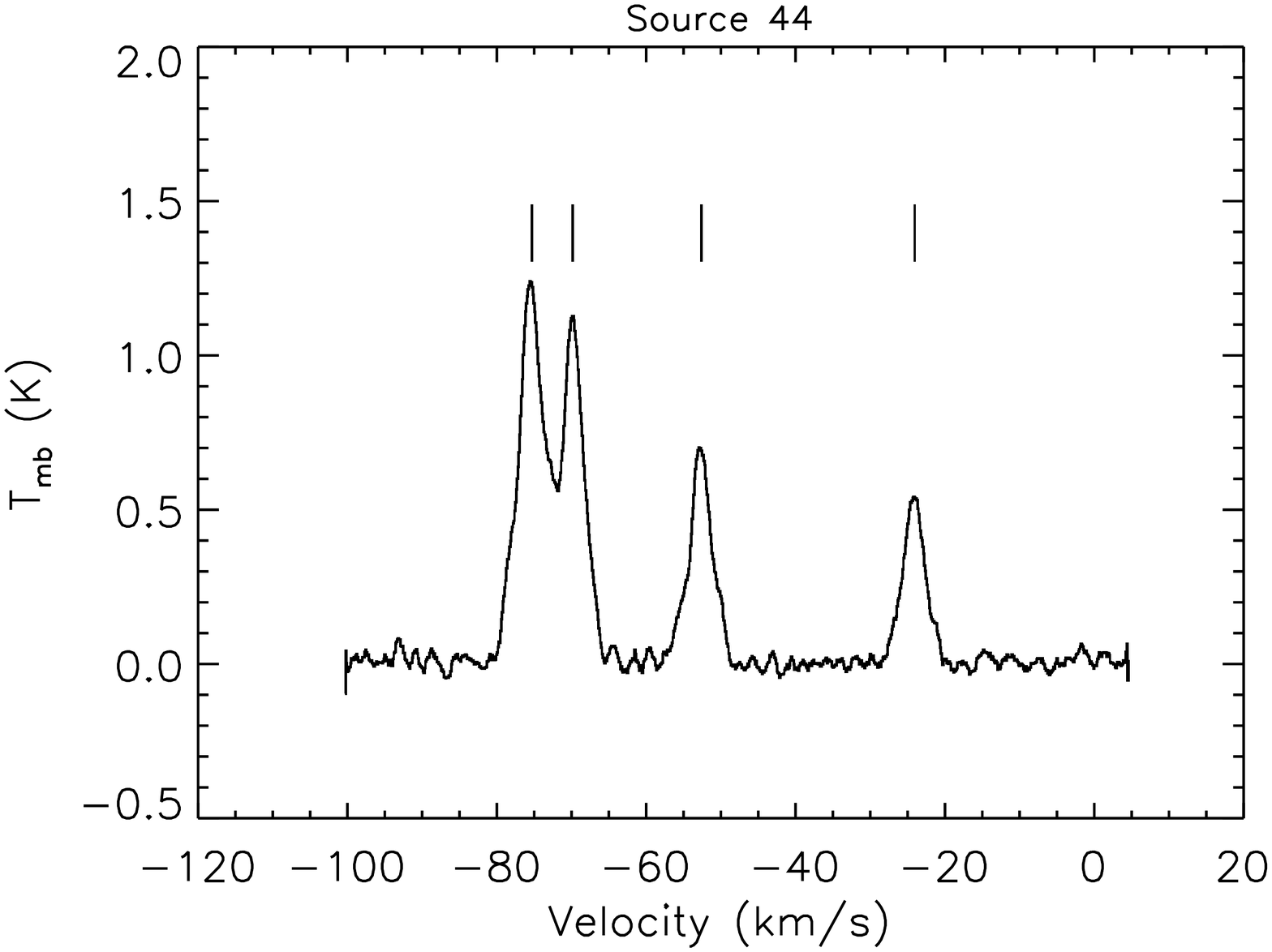}
\includegraphics[trim=1.5cm 1.5cm 1.5cm 1.5cm, width=0.33\textwidth]{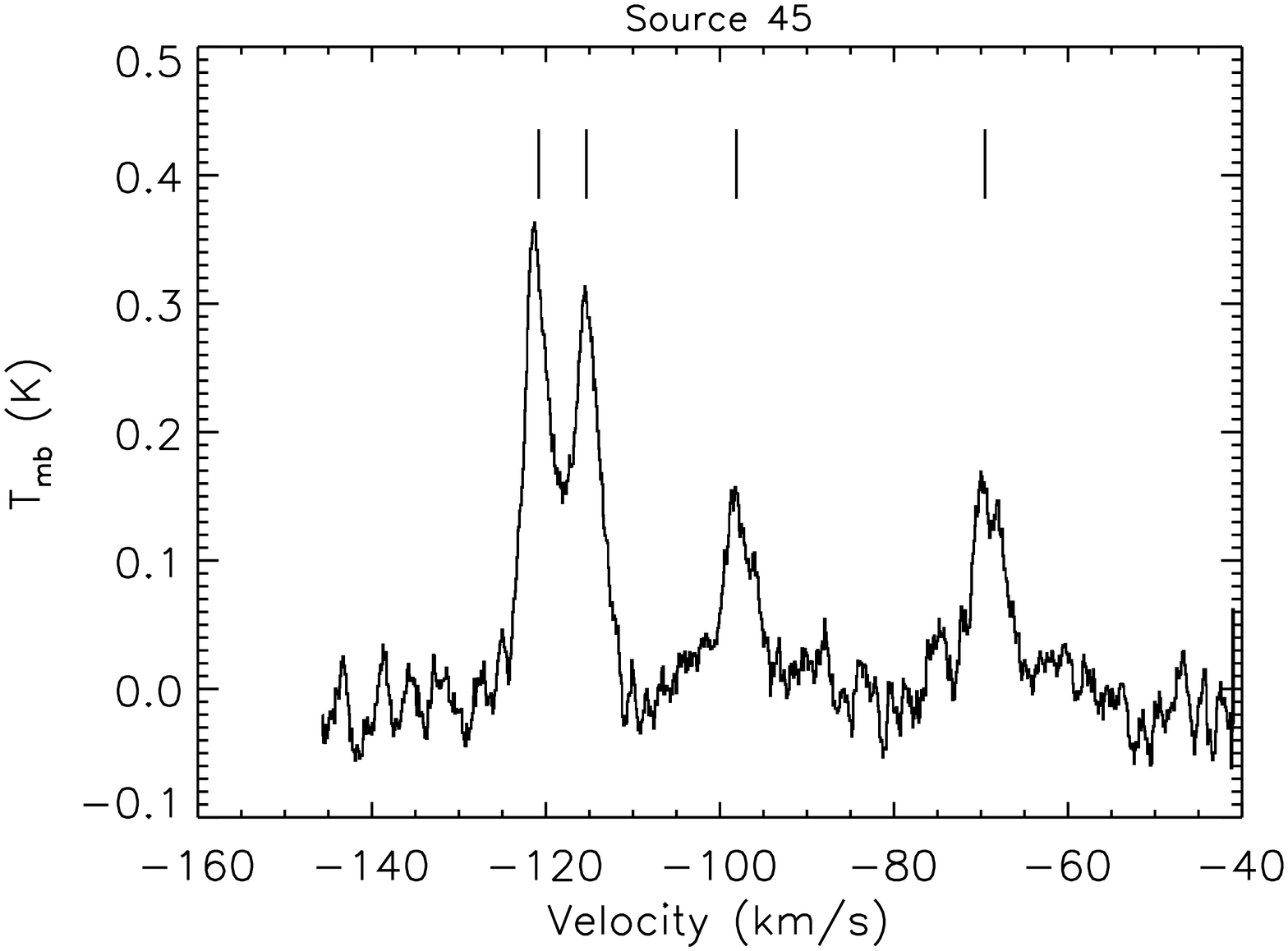}
\includegraphics[trim=1.5cm 1.5cm 1.5cm 1.5cm, width=0.33\textwidth]{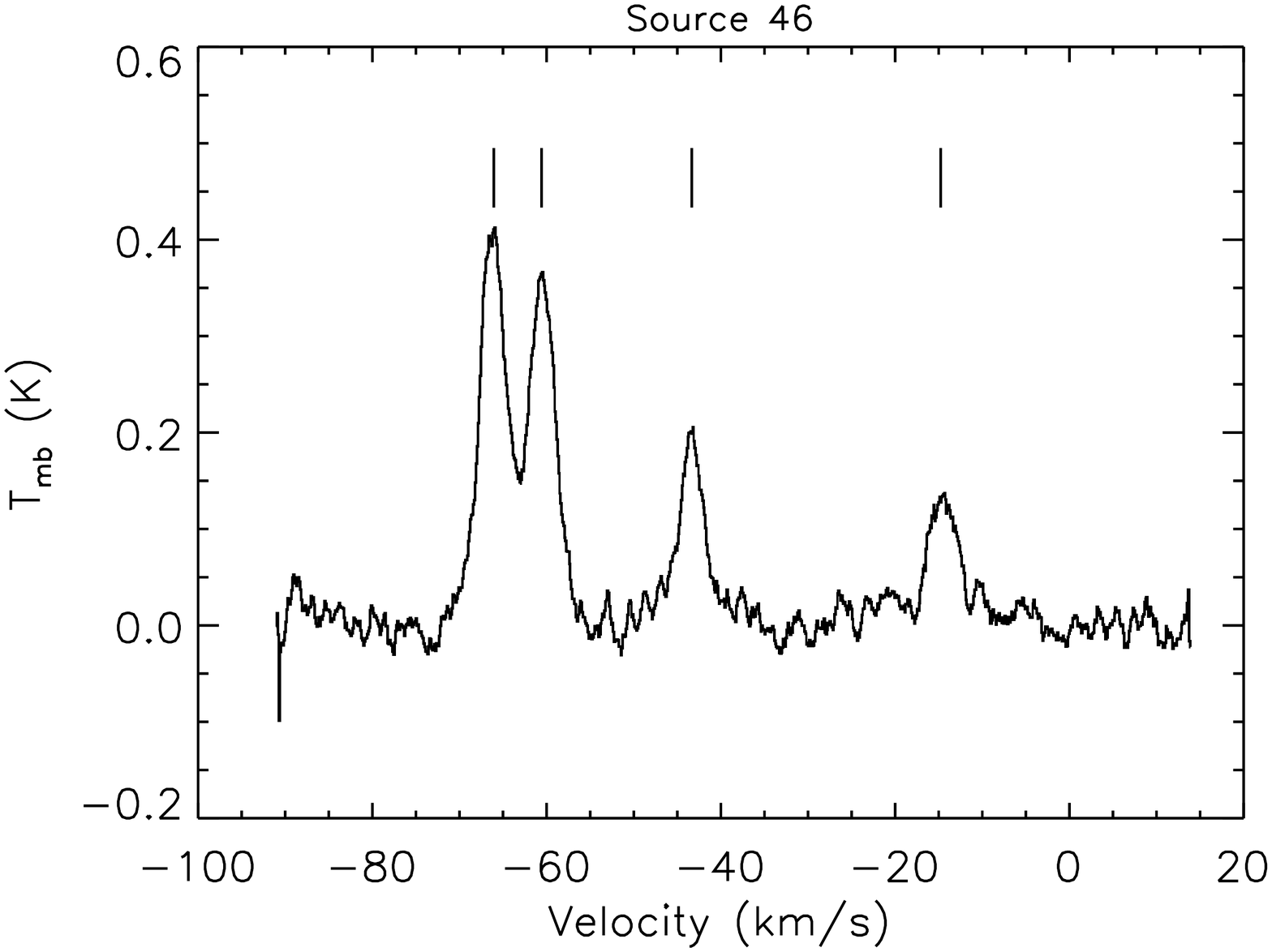}
\includegraphics[trim=1.5cm 1.5cm 1.5cm 1.5cm, width=0.33\textwidth]{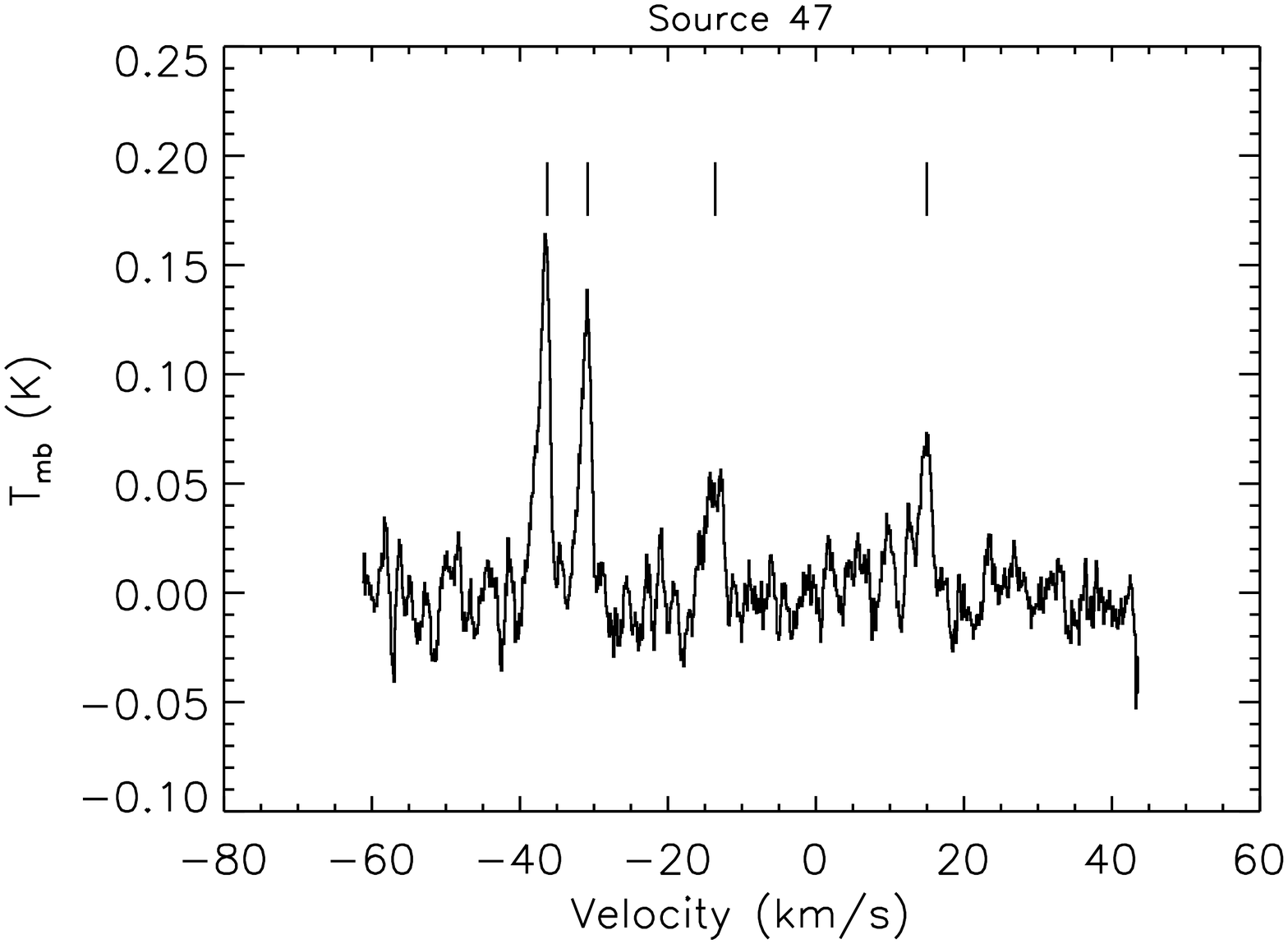}
\includegraphics[trim=1.5cm 1.5cm 1.5cm 1.5cm, width=0.33\textwidth]{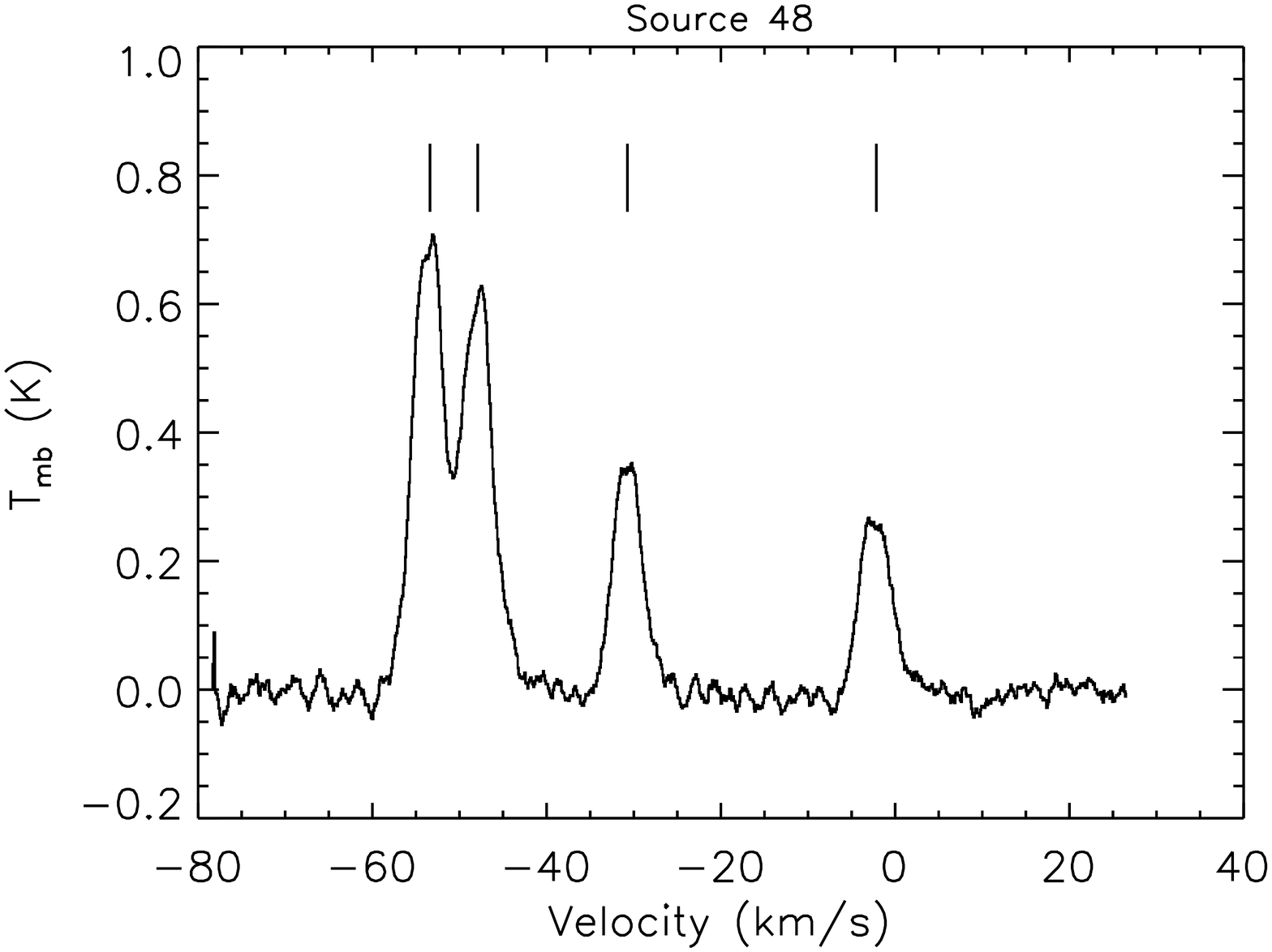}
\includegraphics[trim=1.5cm 1.5cm 1.5cm 1.5cm, width=0.33\textwidth]{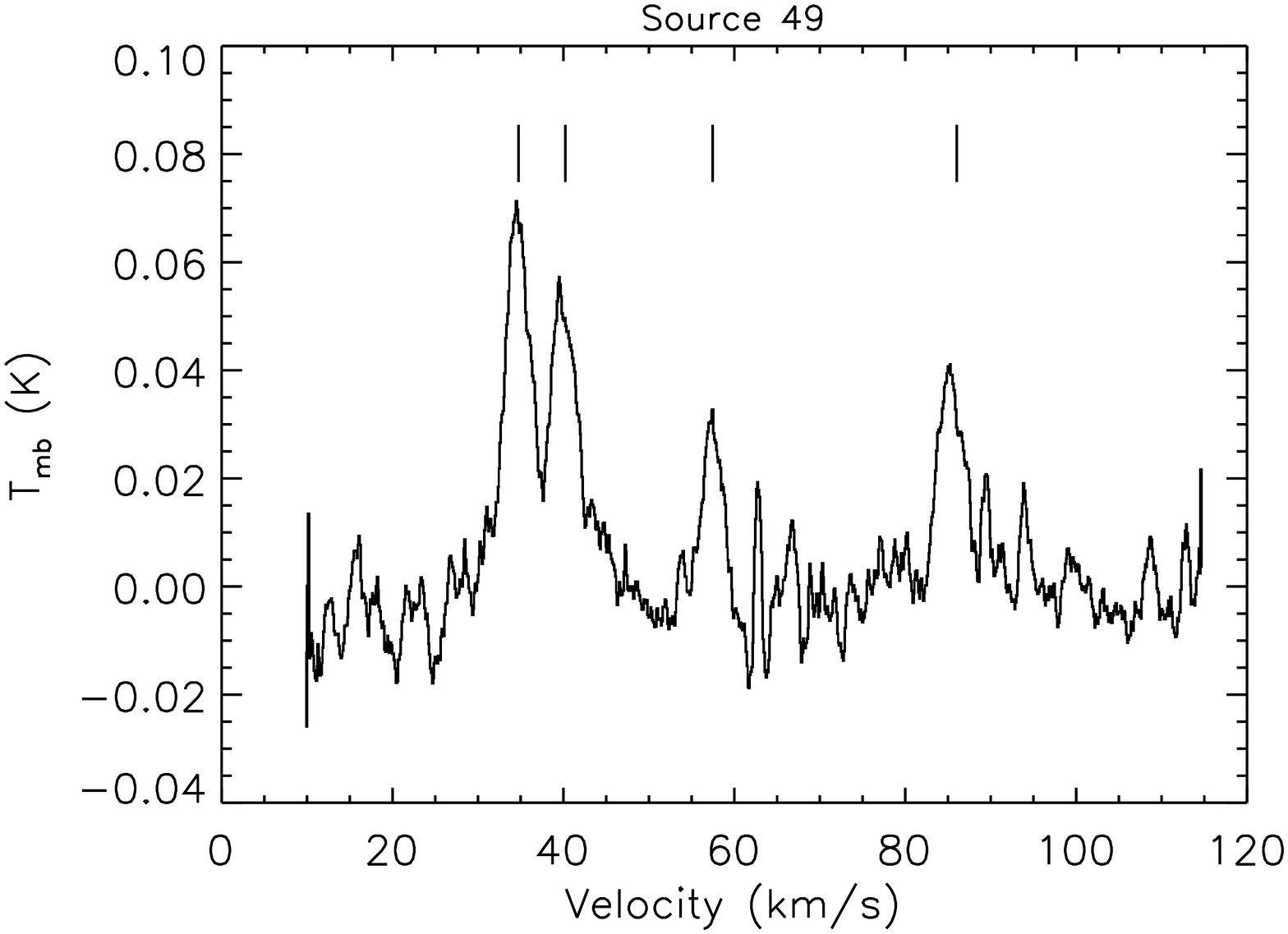}
\includegraphics[trim=1.5cm 1.5cm 1.5cm 1.5cm, width=0.33\textwidth]{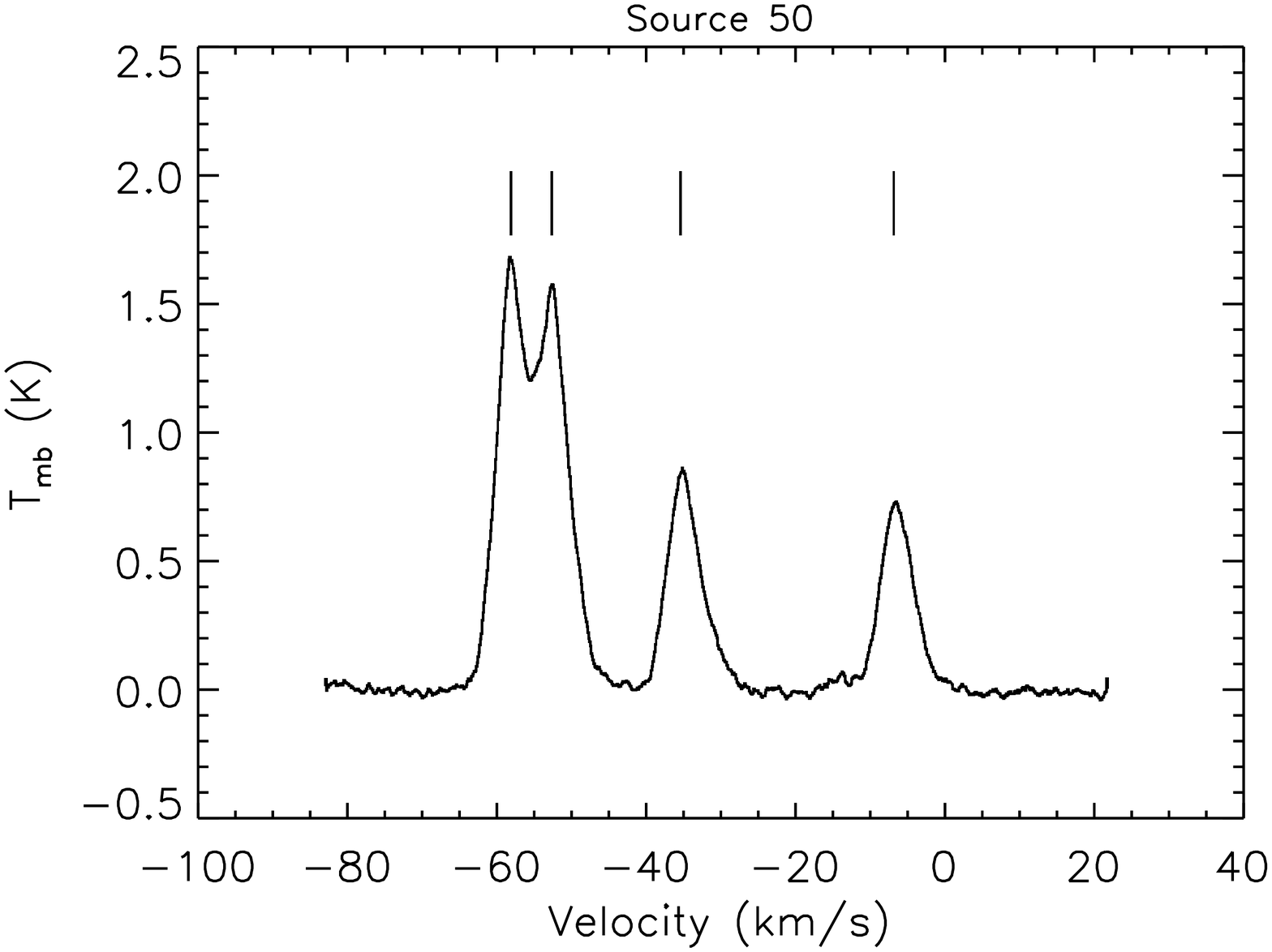}
\includegraphics[trim=1.5cm 1.5cm 1.5cm 1.5cm, width=0.33\textwidth]{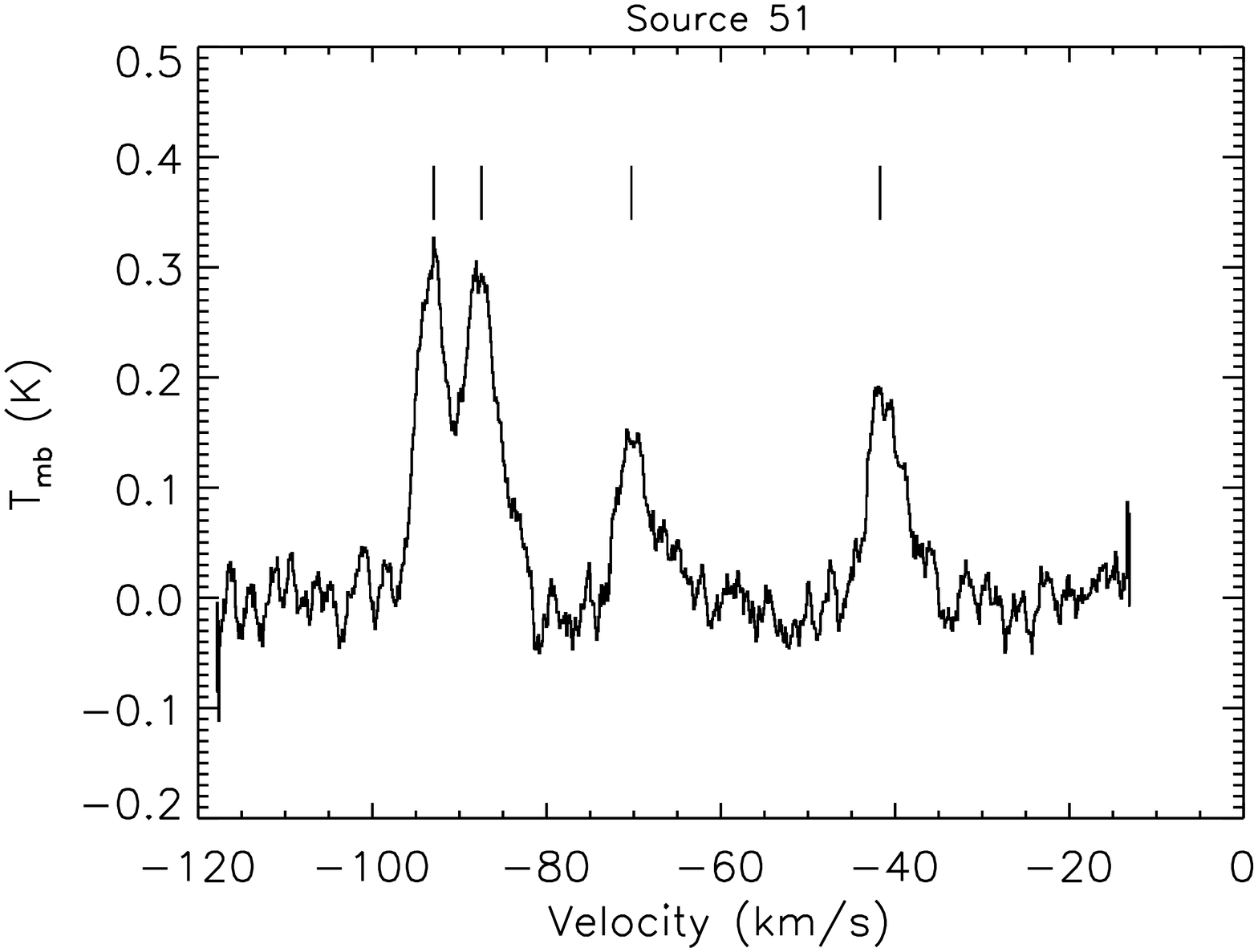}
\caption{Continued.}
\end{figure*}

\section{Results}

\subsection{\ch\ detection statistics}

The sources listed in Table \ref{targetlist} are ordered according to their L/M value computed from the Herschel fluxes, augmented with ancillary data at mid-IR wavelengths \citep{Elia+2016}, and it is immediate to see that the detection rate in \cht\ is strongly dependent on the L/M values. We can distinguish three ranges of L/M: the rate is 0 (none out of 23 sources) for the sources with L/M$< 1$, 58\% (7 out of 12) for sources with 1$<$L/M$<$10, and 90\% (16 out of 18 sources) for sources with L/M$>$10. In three cases, sources \# 32, 35 and 38 as indicated in the table, at least the K=0 line is detected but the 4-component Gaussian fit to the four lines did not converge and so no further analysis could be carried out for these sources. 

This finding is preliminary confirming the working hypothesis that higher values of L/M should be related with a more advanced evolutionary stage. As \ch\ is a temperature probe for gas with densities in excess of 10$^5$ \cmthree\ \citep{Bergin+1994}, its values are influenced more by the inner clump regions and relatively less by the outer envelope. 
Looking at Figure \ref{lm-diag}, the range of L/M \lapprox\ 1 is characteristic of "pre-stellar" clumps (in which, contrary to the sources of the present study, 70 \um\ emission is not detected); the internal input energy producing the detectable 70 \um\ flux would not seem sufficient to drive up considerably the bolometric luminosity as well as the internal clump  temperature, resulting in non-detections in \cht\ for any of the sources with L/M$<$1. This suggests that in these clumps intermediate or high-mass star formation has not yet started. 

Above the L/M=1 threshold, corresponding to the upper envelope of the "pre-stellar" clumps distribution in Fig. \ref{lm-diag}, the \cht\ detection rate increases sharply, implying that the internal energy produced by star formation is rising to levels sufficient to heat up the inner envelope gas to at least $\sim$30 K. The dust temperature estimated using Herschel data \citet{Elia+2016} in these clumps is on average 60\% lower, confirming that the \cht\ line detected is excited in the inner regions of the clumps compared to the more external regions where the $\lambda \geq 100$ \um\ continuum is generated. 

\cite{Miettinen+2006} report a \ch\ abundance that is proportional to temperature in a sample of dense clumps, a result that is interpreted as evidence of desorption from grain ice mantles. The sharp rise in \cht\ detection statistics for L/M\gapprox 1 may then be the result of a combination of the incipient star formation activity in the clumps that raises the temperature at the levels required to excite the (12-11) transition ($E_U$=37K) and to desorb a sufficient amount of \ch\ from the the grains ice mantle. Although a more quantitative characterization of the effect would require mapping data in this species, that we do not have, we believe that the qualitative evolutionary interpretation of the changes in \ch\ detection rate with L/M is convincing.

Above the L/M=10 threshold the energy budget produced internally is such that clumps are all detected in \ch.

\subsection{The \ch\ gas temperature$-$L/M relationship}
\label{lm-t}

\begin{figure}
\includegraphics[width=0.5\textwidth]{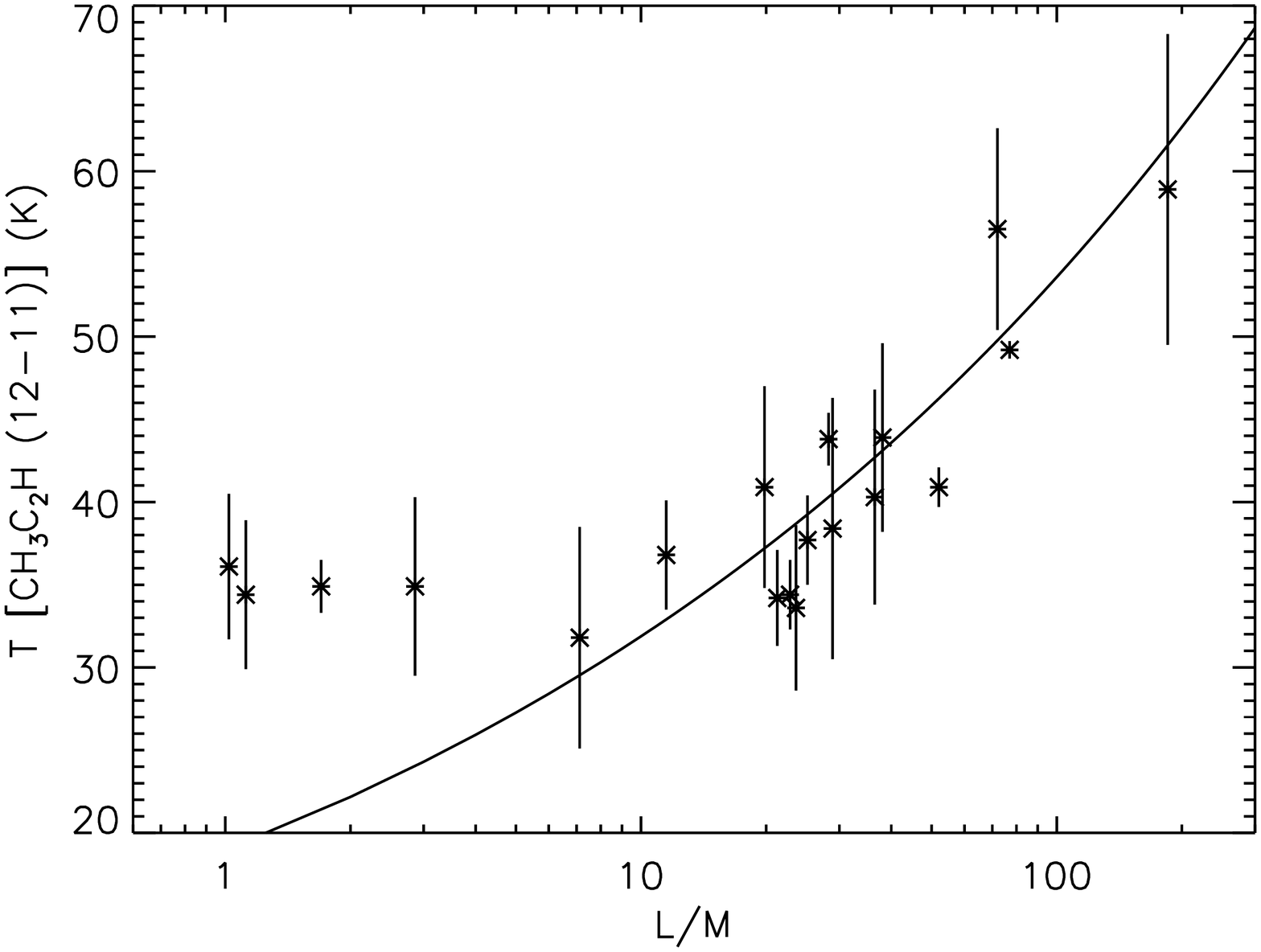}
\caption{\ch\, rotational temperature as a function of the clump L/M. The full line represents the power-law relationship discussed in the text ($T_6 \propto L_{inp} ^{0.22}$ discussed in \S\ref{lm-t}, at constant mass and arbitrarily scaled to fit the L/M\gapprox 10 points). \label{lm-t_plot}}
\end{figure}

While the \cht\ detection statistics therefore confirms the working hypothesis of L/M as an evolutionary indicator, the information derived from the measured temperature of the \ch\ gas provide a clear \emph{quantitative} indication in this sense. Figure \ref{lm-t_plot} presents the \ch\ temperature as a function of L/M, for the detected sources (all with L/M $>$ 1). We find that the derived temperature for all sources detected between L/M=1 and $\sim$10 is T\lapprox\ 35 K irrespectively of L/M. For L/M$>$10 the temperature increases roughly in a linear way with Log(L/M). 

\begin{figure}
\includegraphics[width=0.5\textwidth]{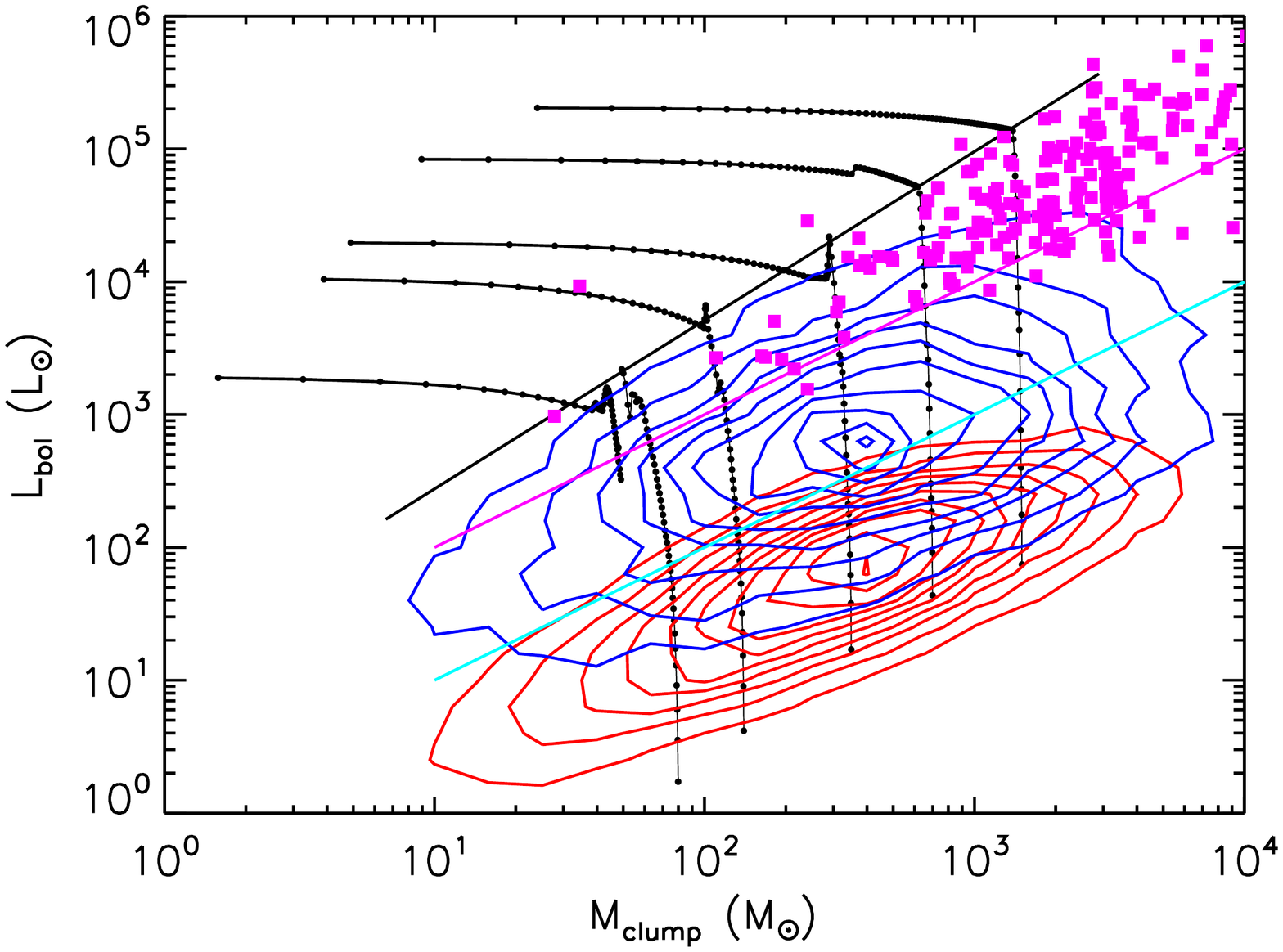}
\caption{Bolometric luminosity $vs$ clump mass for Hi-GAL detected clumps from \cite{Elia+2016}, with distribution of "pre-stellar" clumps (no 70 \um\ counterpart) indicated with red source density contours and distribution of protostellar clumps (with 70 \um\ counterpart) indicated with the  blue contours; tracks are from \cite{Molinari+2008}. Magenta squares are the subsample of protostellar clumps that have a UCHII counterpart in the CORNISH survey \citep{Purcell2013} and that have been studied in detail by \cite{Cesaroni+2015}. The cyan line represents the L/M=1 threshold where \ch\ first detections are found, while the magenta line represent the L/M=10 threshold where T(\ch) starts increasing with Log(L/M) (see Fig. \ref{lm-t_plot}). \label{lm-diag}}
\end{figure}

The L/M thresholds that we indicatively put at 1 and 10 to characterize \ch\ detection statistics seem then to have a more robust foundation in change in the energy budget produced by the star formation process in these clumps, also looking at the location of these two L/M thresholds in the L-M diagram of Figure \ref{lm-diag}: the cyan line for L/M=1 and the magenta line for L/M=10. We already noted how the L/M=1 threshold that marks a abrupt change in \cht\ detection statistics corresponds to the upper envelope of the distribution of "pre-stellar" clumps from \cite{Elia+2016}; the protostellar sources in this L/M range are indistinguishable from "pre-stellar" (other than for the 70\um\ counterpart), and indeed \cht\ is not detected. 

Protostellar clumps leave the L-M area occupied by the "pre-stellar" clumps at L/M=1, where new processes must be kicking in to raise the temperature of the inner regions of the clump to levels that allow \cht\ detection; however, a further moderate increase in L/M up to 10 does not seem to translate into a significant increase of temperature. 
In a  simplified model of a spherical and optically thick dust clump heated by an internal source 
we would  expect that a raise of a factor 10 (for the same clump mass) of the bolometric luminosity should generate a detectable change in temperature. To quantify this effect, we carried out a set of tests with the dust radiative transfer code TRANSPHERE-1D made publicly available by C.P. Dullemond (adapted from \citealt{Dullemond+2002}) to model the clump SED and temperature structure in spherical geometry. 
We initially fixed the inner envelope radius to R$_{int}$=200AU and density gradient $\rho\propto \rho_{int} r^{-1.8}$, varying the internal input luminosities L$_{inp}$ and inner envelope densities. For each run we computed a volume-averaged dust temperature $T_6$ in the region where $\rho \geq 10^6$ \cmthree, that is the gas density regime traced by our observations; we note that dust and gas should be thermally coupled in these dense regimes (e.g., \citealt{Crimier+2010}). We obtain that the two quantities are always related by a power-law of the form $T_6 \propto L_{inp}  ^{0.22}$; the other model parameters (like inner density or outer radius) affect the proportionality constant and are not interesting here as we want to characterise a trend in relative terms.
The $T_6 - L_{inp} $ relationship is reported in Fig. \ref{lm-t_plot} where only the Y-axis scaling has been arbitrarily adjusted to fit the points with L/M\gapprox\ 10; the rise in temperature of such points as a function of L/M is remarkably well-reproduced by the model. 

The few points with L/M$<10$ in the figure, however, do not follow this expected trend. We speculate that a possible reason for this is that the assumption of a central heating source in a spherically symmetric clump might not apply for such clumps. In a competitive accretion paradigm where the initial protostellar seed objects do not form at the center of the hosting clumps, our model assumption would not apply; it is quite plausible that a more shallow spatial distribution of heating sources in the clumps may give rise to a warmer (on average) envelope than in the case of a distribution strongly peaked at the clump center. In other words, we are proposing that the constancy of temperature with luminosity in the 1\lapprox\ L/M\lapprox\ 10 region is not due to a missing increase of temperature as a function of L/M, but rather to an excess temperature for relatively low L/M due to the formation of protostellar seeds away form the clump center. We will investigate this possibility in detail in a forthcoming paper.


Together with the \cht\ detection statistics going virtually to 100\%, the rise in gas temperature in Fig. \ref{lm-t_plot} for L/M\gapprox 10 suggests a further increase of internal input luminosity at the center of the clump that could be due to an increase in the number of forming YSOs in the clump as well as to their increased luminosity production due to a substantial raise in mass. We believe that the second effect is the dominant one. The magenta large squares in Figure \ref{lm-diag} represent the Hi-GAL sources where radio continuum from thermal free-free modelled as HII region emission has been identified by \cite{Cesaroni+2015} in the CORNISH survey \citep{Purcell2013}. The remarkable correspondence of the L/M=10 threshold (indicated with the magenta line in Figure \ref{lm-diag}) with the lower envelope of the distribution of Hi-GAL clumps associated with HII regions, strongly suggests that the rise in gas temperature for L/M\gapprox 10 is mostly due to the appearance of a first ZAMS intermediate/high-mass star in the clump. 

Based on the results presented we propose an evolutionary classification for massive protostellar clumps in which a 70 \um\ counterpart is detected, based on the L/M ratio of the clumps: 

\begin{itemize}

\item A relatively quiescent phase characterised by L/M\lapprox 1 where relatively low-mass objects may be forming. Such seeds of star formation might be either low-mass YSOs or intermediate/massive YSOs in the pristine stages of evolution. These clumps differ from pre-stellar clumps of similar L/M in the appearance of a 70\um\ counterpart detection; yet the internal input energy does not seem sufficient to warm up the clumps to a detectable level in \ch.


\item An intermediate phase where 1\lapprox L/M \lapprox 10, in which the clump internal input power due to star formation is sufficient to warm up the envelope and be detectable in \ch, but where the increase in input power does not seem to reflect in a temperature increase of the densest portions of the envelope, as expected from simple radiative transfer models of spherical clumps heated by a central power source. We argue that this could reflect a phase, consistently with protocluster competitive accretion scenarios, in which protostellar seeds are not forming strictly in the center of the clumps but with a more shallow spatial distribution in the clumps; plausibly such a configuration, for which detailed radiative transfer modelling is needed, is more efficient to warm up the inner regions of the envelope than the centrally peaked input power case. While the luminosity of forming protostars increases, their migration toward denser regions of the clumps makes them less effective in raising the overall envelope temperature.


\item A third phase with L/M\gapprox 10, where the further dramatic increase in luminosity starts warm up the clump more and more. The association of the L/M=10 threshold with the location in the L-M diagram where HII region counterparts to Hi-GAL sources are found (see Figure \ref{lm-diag}) is an indication that this luminosity increase is related to the birth of the first intermediate/high-mass ZAMS stars in the clusters.

\end{itemize}

In a parallel effort we are producing an extensive Montecarlo grid of SED models for synthetic protoclusters embedded in massive clumps spanning masses between 10$^2$ and 10$^4$ \msun, dust temperatures between 10 and 40K, ages of embedded YSOs cluster from 5\,10$^4$ to 5\, 10$^5$ years and star formation efficiencies up to 40\%  (Molinari \& Robitaille, in preparation). A preliminary analysis of the properties of the models grid shows that irrespectively of the various models parameters, 
$\sim$80\% of the models showing L/M$\geq$10 contain at least one ZAMS star and, for clump masses greater than 10$^3$\msun\ (our source selection criterion), the most massive ZAMS star formed is of spectral class B1 or earlier for nearly 100\% of the models. A thorough analysis and discussion is deferred to the paper in preparation, but this preliminary analysis seem to confirm that the birth of a B1 ZAMS star (or earlier) is associated with the crossing of the L/M=10 threshold; such stars nominally produce N$_{Ly} \sim 45.11$ \citep{thompson84} that is comparable to the lower values of the Lyman continuum derived by \cite{Cesaroni+2015} from the radio flux of the CORNISH counterparts to Hi-GAL clumps. 

\section{Conclusions}

Observations of the \cht\ line towards 51 protostellar massive clumps selected from the Hi-GAL survey provide a clear indication that the ratio between the bolometric luminosity and the mass of a clump can be used to diagnose the star formation evolutionary stage of the clumps. Three stages are identified corresponding to three intervals in L/M values. L/M\lapprox 1 in which star formation is either in its very early stages or only low-mass YSOs are forming; 1 \lapprox L/M \lapprox 10 in which clumps build up their luminosities and temperature, due to ongoing evolution of relatively low-mass protostars; L/M \gapprox 10 in which inner clump gas temperature rises with L/M likely due to the first appearance of intermediate and high-mass ZAMS stars.

\acknowledgments

This work is based on observations collected at the European Southern Observatory under programmes 096.C-0920(A) and 296.C-5011(A), and is part of the VIALACTEA Project, a Collaborative Project under Framework Programme 7 of the European Union funded under Contract \# 607380,  that is hereby acknowledged. The work was also partially supported by the Italian Ministero dell'Istruzione, Universit\`a e Ricerca, through the grant "Progetti Premiali 2012 - iALMA" (CUP CS2I3000140001).


We thank C. de Breuck and the APEX staff for supporting the APEX observations, and V. Rivilla for illuminating discussions on the chemistry of \ch.

\end{document}